\documentclass{vldb1}       
\usepackage{epsfig}
\usepackage{graphicx}
\usepackage{amsmath}
\usepackage{url}
\usepackage{pseudocode}
\usepackage{txfonts}
\usepackage{tipa}
\usepackage{caption}
\usepackage{subfigure}
\usepackage{balance}  
\usepackage{epsfig}
\usepackage{amssymb}
\usepackage{amsfonts}
\usepackage{times}
\usepackage{xcolor}
\usepackage[boxed]{algorithm2e}
\usepackage[latin1]{inputenc}
\usepackage{tabularx}
\usepackage{multirow}
\usepackage{epstopdf}
\DeclareGraphicsExtensions{.eps}

\usepackage{pifont}
\DeclareMathSizes{10}{10}{10}{10}
\newcommand{\argmax}{\arg\!\max}
\usepackage{color}

\newcommand{\laure}[1]{\textcolor{red}{#1}}
\makeatletter
\newlength{\earraycolsep}
\def\eqnarray{\stepcounter{equation}\let\@currentlabel%
\theequation
\global\@eqnswtrue\m@th
\global\@eqcnt\z@\tabskip\@centering\let\\\@eqncr
$$\halign to\displaywidth\bgroup\@eqnsel\hskip\@centering
$\displaystyle\tabskip\z@{##}$&\global\@eqcnt\@ne
\hskip 2\earraycolsep \hfil$\displaystyle{##}$\hfil
&\global\@eqcnt\tw@ \hskip 2\earraycolsep
$\displaystyle\tabskip\z@{##}$\hfil
\tabskip\@centering&\llap{##}\tabskip\z@\cr}
\makeatother

\makeatletter    
\newcommand{\unnumberedcaption}%
	{\@dblarg{\@unnumberedcaption\@captype}}
\newcommand{\@unnumberedcaption}{}
\long\def\@unnumberedcaption#1[#2]#3{\par
  \addcontentsline{\csname ext@#1\endcsname}{#1}{%
    \protect\numberline{}{\ignorespaces #2}%
    }%
  \begingroup
    \@parboxrestore
    \normalsize
    \@makeunnumberedcaption{\ignorespaces #3}\par
  \endgroup}
\newcommand{\@makeunnumberedcaption}[1]{%
  \vskip\abovecaptionskip
  \sbox\@tempboxa{#1}%
  \ifdim \wd\@tempboxa >\hsize
    #1\par
  \else
    \global \@minipagefalse
    \hbox to\hsize{\hfil\box\@tempboxa\hfil}%
  \fi
  \vskip\belowcaptionskip}
\@ifundefined{abovecaptionskip}{%
  \newlength{\abovecaptionskip}%
  \setlength{\abovecaptionskip}{10pt}%
}{}
\@ifundefined{belowcaptionskip}{%
  \newlength{\belowcaptionskip}%
  \setlength{\belowcaptionskip}{0pt}%
}{}

\makeatother    
\renewcommand{\S}{{\cal S}}

\begin{document}
\sloppy


\title{ 
Truth Discovery Algorithms: An Experimental Evaluation\\
QCRI Technical Report, May 2014}
\numberofauthors{2} 
\author{
\alignauthor Dalia Attia Waguih \\
       \affaddr{Qatar Computing Research Institute}\\
       \affaddr{Doha, Qatar}\\
       \email{dattia@qf.org.qa}
\alignauthor Laure Berti-\'Equille\\
       \affaddr{Qatar Computing Research Institute}\\
       \affaddr{Doha, Qatar}\\
       \email{lberti@qf.org.qa}
}\maketitle

\vspace{-.3cm}\begin{abstract}
A fundamental problem in data fusion is to determine the veracity of multi-source data in order to resolve conflicts. 
While previous work in truth discovery  has proved to be useful in practice for specific settings, sources' behavior or data set characteristics, there has been limited systematic comparison of the competing methods in terms of efficiency, usability, and repeatability. 
We remedy this deficit by providing a comprehensive review of 12 state-of-the art algorithms for truth discovery. We provide reference implementations and an in-depth evaluation of the methods based on extensive experiments on synthetic and real-world data. 
 We analyze aspects of the problem that have not been explicitly studied before, such as the impact of  initialization and parameter setting, convergence, and scalability. We provide an experimental framework for extensively comparing the methods in a wide range of truth discovery scenarios where  source coverage, numbers and distributions of conflicts, and true positive claims can be controlled and used to evaluate the quality and performance of the algorithms. Finally, we report comprehensive findings obtained from the experiments and provide new insights for future research.
 
\end{abstract}

\section{Introduction}

\begin{table*}
\begin{minipage}{.6\textwidth}
\centering\scriptsize
\begin{tabular}{r|p{.1cm}|p{.6cm}|p{.65cm}|p{.65cm}|p{.65cm}||p{.65cm}|p{.6cm}|p{.65cm}|}
\cline{3-6}
 \multicolumn{2}{l|}{{\bf (a)  Affiliations}}	& $S_1$  	& $S_2$				&  $S_3$	& $S_4$		&\multicolumn{1}{c}{ {\scriptsize\texttt{GT}}	}&	\multicolumn{1}{c}{{\scriptsize\texttt{Conf}}}&\multicolumn{1}{c}{} \\ 
\cline{2-8}   $d_1$&{\it Stonebraker}& MIT 		& UWisc				&  -			& MIT			&	\multicolumn{1}{c}{MIT	}&	\multicolumn{1}{|c}{2}&\multicolumn{1}{|c}{} \\ 
\cline{2-8}  $d_2$&{\it Bernstein} 	& MSR 		& - 					& AT\&T		&-				&	\multicolumn{1}{c}{MSR}		&\multicolumn{1}{|c}{2}&\multicolumn{1}{|c}{} \\ 
\cline{2-8}  $d_3$&{\it Carey} 			& UCI  		& -						& BEA			& BEA			&	\multicolumn{1}{c}{UCI}	&	\multicolumn{1}{|c}{2}&\multicolumn{1}{|c}{} \\ 
\cline{2-8}  $d_4$&{\it Halevy} 			& Google	& -						& UWisc			& MSR			&\multicolumn{1}{c}{	Google	}	&\multicolumn{1}{|c}{3}&\multicolumn{1}{|c}{} \\ 
\cline{2-8}
\cline{2-7}  \multicolumn{2}{c}{ {\scriptsize\texttt{Cov }} }& \multicolumn{1}{c}{1	}& \multicolumn{1}{c}{.25	}& \multicolumn{1}{c}{.75}	& \multicolumn{1}{c}{.75}& \multicolumn{3}{c}{}\\

\multicolumn{9}{l}{}\\
 \cline{3-6}
\multicolumn{2}{l|}{{\bf (b)  Src Truthworthiness}}&$T_{S_1}$  	& $T_{S_2}$			&  $T_{S_3}$	& $T_{S_4}$	&	\multicolumn{3}{c}{	{\bf Algorithm Precision}}\\ 
\cline{2-9}
& {\scshape Depen}&	0.0323	&0.0253&0.0297&0.0382&	\multicolumn{3}{c|}{.75} \\
\cline{2-9}
& {\scshape TruthFinder}&		0.0489&0.0489&0.0489&0.0489&	\multicolumn{3}{c|}{.25}\\
\cline{2-9}
\multicolumn{9}{c}{}\\
\multicolumn{9}{l}{{\bf (c) Value Confidence by  {\scshape TruthFinder}}	}\\ 
\cline{3-9}
\multicolumn{2}{l}{}&\multicolumn{1}{|c|}{MIT}&MSR&UWisc&\multicolumn{1}{c|}{BEA}&Google&UCI&AT\&T\\
\cline{2-9}  $d_1$&{\it Stonebraker}&{\bf 0.5025}	& &0.5009& \multicolumn{1}{c|}{}&&&\\
\cline{2-9}  $d_2$&{\it Bernstein} 	& 						&0.50100&&\multicolumn{1}{c|}{}&&&{\bf  0.50101}\\
\cline{2-9}  $d_3$&{\it Carey} 			 			&&		&   &\multicolumn{1}{c|}{{\bf  0.5024}}&&0.5007&\\
\cline{2-9}  $d_4$&{\it Halevy} 		&							&{\bf 0.50071} & 0.50067&\multicolumn{1}{c|}{}&\multicolumn{1}{c|}{0.50065}&&\\
\cline{2-9}
\end{tabular}
\caption*{{\bf Table 1. Illustrative Example}}\label{tab:example} 
\end{minipage}
\hfill\
\begin{minipage}{.37\textwidth}
\centering\scriptsize
\begin{tabular}{|l|l|}
\hline
\multicolumn{2}{|c|}{ {\bf Notation}  } \\ 
\hline $S$ & Set of all sources \\ 
\hline $S_v$ & Set of sources providing value $v$\\
\hline $S_{\bar{v}}$ & Set of sources providing a distinct value from $v$ \\
\hline $D$ & Set of data items as (object, attribute) pairs \\ 
\hline $D_s$ & Set of data items covered  by source $s$\\ 
\hline $D_v$ & Set of data items corresponding to value $v$ \\ 
\hline $V_d$ & Set of values provided for  data item $d$\\
\hline $V$ & Set of all values for all data items \\ 
\hline $V_{D_s}$& Set of values for the data items provided by source $s$\\
\hline $V_s$ & Set of values provided by source $s$ \\ 
\hline $T_s$ & Truthworthiness of source $s$\\
\hline $C_v$ & Confidence of value $v$\\
\hline 
\end{tabular}
\caption*{{\bf Table 2. Notations}}\label{tab:notations}
  \end{minipage}
\vspace{-.5cm}
\end{table*}

As online user-generated content grows exponentially, the reliance on Web data 
is inevitably growing in every application domain. 
However, data can be biased, noisy, 
outdated, incorrect, and thus, misleading and unreliable. Massive data coming from multiple sources 
amplifies the difficulty  of  ascertaining information veracity. 
The problem of truth discovery is  intellectually 
and technically interesting enough to have attracted a lot of prior studies, from  the artificial intelligence and the database communities, 
 sometimes investigated under the names of  fact-checking~\cite{GoasdoueKKLMZ13}, information credibility~\cite{PasternackR13}, information corroboration~\cite{GallandAMS10}, data fusion ~\cite{Pochampally2014,sigmod14}, conflicting data integration~\cite{DongBS09}, or knowledge fusion~\cite{vldb14}. Truth discovery problem can be formulated as follows. Given a set of assertions claimed by multiple sources, label each claimed value as true or false and compute the reliability of each source. One major line of  work  extends truth discovery
models by incorporating prior knowledge either about the claimed assertions (e.g., {\scshape SimpleLCA} and {\scshape GuessLCA} \cite{PasternackR13}) or about the source 
reputation via trust assessment (e.g., SourceRank \cite{BalakrishnanK11}). Another important line of research aims at iteratively computing 
and updating the trustworthiness of a source as a function of the belief in its claims, and then the belief score of each 
claim as a function of the trustworthiness of the 
sources asserting it (e.g., {\scshape TruthFinder} \cite{YinHY08}). In this line, several probabilistic models have been proposed to incorporate various aspects beyond source 
trustworthiness and claim belief, namely: the dependence between sources (e.g., {\scshape Depen} and its extensions \cite{DongBHS10a}), the temporal 
dimension in discovering evolving truth  \cite{DongBS09a}, the difficulty of ascertaining the veracity of certain claims (e.g., {\scshape Cosine}, 2- and {\scshape 3-Estimates} \cite{GallandAMS10}), and the management of collections of entities (e.g., LTM~\cite{ZhaoRGH12}) or linked data \cite{GoasdoueKKLMZ13}.

 There are a number of challenges in truth discovery.  The first challenge is a theoretical one since it is difficult to formalize a method general enough to handle various data set characteristics and truth discovery scenarios. We observe that none of the methods constantly outperforms the others in terms of precision and a ``one-fits-all'' approach does not seem to be achievable.  
  Another challenge is related to the usability of the methods.  Assumptions of  truth discovery models and complex parameter setting make current approaches still difficult to use and apply to the  wide diversity of information available on the Web. 
   
  {\bf Related Work.}  Previous comparative studies such as the work of Li  {\it et al.} \cite{LiDLMS12} and \cite{sigmod14} are based on real-world data sets and {\it gold standards} because, in practice, the {\it complete ground truth} often does not exist or is out-of-reach. Such gold standards are samples of the ground truth (generally less than 10\% of the original data set's size). We claim that they are not statistically significant to be legitimately used for evaluating and comparing existing methods in a systematic way. Moreover, previous comparisons did not study important algorithmic aspects of the methods such as parameter settings, time complexity, repeatability, computational issues, scalability, and convergence of the algorithms. They did not test them extensively for a wide range of truth discovery scenarios systematically generated with the control of the {\it complete} ground truth distribution. The experimental framework and data set generator we propose for comparing the methods are novel, practical contributions to the field,  so that others can use and extend them for benchmarking, parameter setting and tuning of  existing and new truth discovery algorithms. Publicly-available data sets with complete ground truth are notoriously difficult to obtain. The data set generator can serve as a useful proxy for what-if scenarios and reproducibility, to understand, in a systematic way, the data set characteristics that have significant impact on the performance and quality of the algorithms.  
 The goals of our study are:\\
(1) To provide a clear explanation of each algorithm
, and allow comparison of their properties by using common notation, terminology,  experimental set-ups,  data sets, and test cases, \\
(2) To provide reference implementations of these algorithms
against which future algorithms can be compared, new data sets can be analyzed, 
and on top of which algorithms for different problems or applications can be
built, and finally, \\
(3) To perform a thorough experimental evaluation of the algorithms
over a variety of data sets and report their performance and quality for a wide spectrum of parameter settings.

This paper is structured as follows. 
In Section 2, we define the problem of truth discovery and 
 describe  the algorithms in detail. 
In Section 3, we present our comparative study based on synthetic data sets systematically generated to demonstrate the quality of the algorithms in various truth discovery scenarios. Then, we study scalability, and finally, we evaluate the methods on five real-world data sets. In Section 4, we recapitulate our findings and conclude the paper.

\section{Truth Discovery Algorithms}
  We consider the truth discovery algorithms that take, as input data, a set of claims in the form of quadruplets $(claimID,sourceID,dataItemID,value)$ and infer, as output result, a Boolean truth label for each claim.  
  In addition, the truth discovery algorithms may also return $T_s$, the truthworthiness of each source $s$, and $C_v$, the confidence of each value $v$. For example, consider the four sources in the example of Table~1(a) adapted from \cite{DongBS09}. They provide claims on affiliation of four researchers such as~{\scriptsize\texttt{(c1,S1,Stonebraker:AffiliatedTo,MIT)}}.  Source coverage {\small\texttt{(Cov)}}  is 1 for $S_1$, .25  for $S_2$, and .75 for $S_3$ and $S_4$. 
  Only $S_1$ actually provides a correct value for each data item, from $d_1$ to $d_4$, in conformance with the ground truth {\small\texttt{(GT)}}. Depending on the number of distinct values per data item  {\small\texttt{(Conf)}} -- e.g.,  $d_1$-$d_3$ have 2 distinct values -- some algorithms can make random guessing or wrong decisions if some sources copy claims from another source. In Table~1(b),  source truthworthiness  has been computed by  {\scshape  Depen} and  {\scshape  TruthFinder} algorithms. The precision is computed from the number of true positives in {\small\texttt{(GT)}} also returned by the algorithms (.75 and .25, respectively). Truthworthiness of $S_1$ is .0489 for {\scshape TruthFinder}, whereas it is .0323 for  {\scshape Depen}. Table 1(c)  shows the confidence of each value computed by {\scshape  TruthFinder}. The values considered to be true by this algorithm are in bold. As illustrated by this example, truth discovery algorithms may have different precision and output results depending on  parameter setting and data set characteristics.  In this paper, we study the effect of both 
    on the quality and performance of 12 truth discovery algorithms from the literature. We use the notations presented in Table~2. 
  Each truth discovery algorithm is presented in detail with its pseudocode where \ding{182} refers to  the computation of value confidence $C_v$, and  \ding{183} refers to the computation of source truthworthiness, $T_S$. We study the impact of various parameter settings on the quality of each algorithm  and we analyze time complexity in Table~3. 
  We made several choices for the consistency and fairness of  our study. First, we initialized source truthworthiness $T_S$ to  .8 for all algorithms because it maximizes the precision of most algorithms. Second, we use the Book data set for this preliminary parameterization study. The Book data set has been formatted in different versions so that all algorithms can be compared from the same input data set. Third, we  use  the same convergence test for all algorithms: the difference of source truthworthiness cosine similarity between two successive iterations to be less than or equal to a given threshold, $\delta$, as we will describe in this section.  We will discuss these choices at the end of the section and conclude on this first set of experiments dedicated to parameter setting. Due to the space limitation, we had to limit the presentation of our results but we invite the reader to access the full set of the experimental results and  codes in \cite{BW2014}.
\begin{table*}
\centering
\small
\scalebox{1}{
\begin{tabular}{p{5.5cm}p{5.5cm}p{5.5cm}}
\begin{pseudocode}{{\scshape TruthFinder}}{S,D,V,\rho ,\gamma, \delta} 
{\bf Initialization. }\\
\forall  s \in S : T_s \gets 0.8\\
\REPEAT
	\FOREACH d\in D
		\DO
				\FOREACH v \in  V_d: \\
					\BEGIN	
						\sigma_v \gets -\sum\limits_{ s\in S_v}  \ln(1-T_s)\\
				
						\sigma^{\star}_v \gets \sigma_v + \rho \sum\limits_{ v'\in V_d}\sigma_{v'} . sim(v,v') \\
					C_v \gets 1/(1+e^{-\gamma\sigma^{\star}_v})$\ding{182}$ 
						\\
					\END\\
	\FOREACH s \in S: T_s \gets \sum\limits_{ v \in V_s}C_v/|V_s|$\ding{183}$ \\
\UNTIL Convergence(T_S,\delta)\\
\FOREACH  d \in D \DO
trueValue(d) \gets \displaystyle\argmax_{ v \in V_d} (C_v)\\
\hrulefill
\end{pseudocode}
\vspace{-0.2in}\begin{pseudocode}{{\scshape Cosine}}{S,D,V,\eta, \delta}
{\bf Initialization. }\hspace{.5cm}$i=1$	\\
\forall s \in  S: 
	\BEGIN 
		T^0_s \gets (2|V_s|-|V_{D_S}|) /|V_{D_S}|\\
		\forall v \in  V_s : C_v \gets 1 
	\END\\
\REPEAT
	\FOREACH s \in  S \DO
		\BEGIN
			pos \gets \sum\limits_{ v \in V_s} C_v ;
			neg \gets \sum\limits_{ v \in V_{D_s} - V_S} C_v\\
			norm \gets (|V_{D_s}|  \sum\limits_{ v \in  V_{ D_s}} C_v^2)^{1/2}\\
			T^i_s \gets (1-\eta)T^{i-1}_s + \eta \frac{pos - neg}{norm}$\ding{183}$\\
		\END\\
	\FOREACH d\in  D \DO
		\FOREACH v \in  V_d \DO
			\BEGIN
				pos \gets \sum\limits_{ s \in S_v}(T^i_s)^3; 
				neg \gets \sum\limits_{ s\in S_d \wedge s \notin S_v}(T^i_s)^3\\
				norm \gets \sum\limits_{ s\in S_d} (T^i_s)^3\\
				C_v \gets \frac{pos - neg}{norm}$\ding{182}$
			\END\\
			$i++$\\
\UNTIL Convergence(T^i_s,T^{i-1}_s,\delta)\\
\FOREACH d \in D \DO
 trueValue(d) \gets   \displaystyle\argmax_{ v \in V_d} (C_v)\\
\end{pseudocode}
&
\begin{pseudocode}{{\scshape 2-Estimates}}{S,D,V,\lambda,\delta}
{\bf Initialization. }\\
\forall s \in  S : T_s \gets 0.8\\
\REPEAT
	\FOREACH d \in  D \DO
		\BEGIN
			\FOREACH v \in  V_d  \DO
				\BEGIN
					pos \gets \sum\limits_{ s \in S_v}(1-T_s)\\
					\\
					 	neg \gets \sum\limits_{s\in S_{\bar{v}}} T_s\\
					 	\\
					C_v \gets \frac{pos + neg}{|S_d|}$\ding{182}$ \\
				\END\\
		\END\\	
List(C_v) \gets Normalize(List(\{C_v|\forall v\}),\lambda)\\
	\FOREACH s \in  S  \DO
		\BEGIN
			pos \gets \sum\limits_{v \in V_s} (1-C_v)\\
			\\
		 	neg \gets \sum\limits_{s\in S_{\bar{v}}} C_v\\
		 	\\
			T_s \gets  \frac{pos + neg}{|V_{D_S}|}$\ding{183}$\\
		\END\\
List(T_s) \gets Normalize(List(\{T_s|\forall s\}),\lambda)\\
\UNTIL Convergence(T_s,\delta)\\
\FOREACH d\in D \DO
  	trueValue(D) \gets   \displaystyle\argmax_{ v \in V_d} (C_v) \\
\hrulefill\\
\\
{\bf Function }\hspace{.2cm}Normalize(List(X),\lambda) \\
\label{alg:normalize}
minX \gets \displaystyle\min(List(X))\\
\\
maxX \gets  \displaystyle\max(List(X))\\
\\
        \FOREACH x \in  List(X)  \DO
            \BEGIN
                x1 \gets \frac{x - minX}{maxX - minX}\\
                \\
                x2 \gets round(x)\\
                \\
                x \gets \lambda . x1 +  (1 - \lambda) x2
            \END\\
            {\bf Return }\hspace{.1cm} List(X)
\end{pseudocode}%
&
\begin{pseudocode}{{\scshape 3-Estimates}}{S,D,V,\lambda,\delta}
{\bf Initialization.}\\ 
\forall s \in  S : T_s \gets 0.8\\ 
\forall d\in  d, \forall v \in  V_d : \varepsilon_v \gets 0.1\\
\REPEAT
	\FOREACH d\in  D \DO
		\BEGIN
			\FOREACH v \in  V_d \DO
				\BEGIN
					pos \gets \sum\limits_{ s \in S_v}(1- T_s\varepsilon_v)\\
					\\
			   	neg \gets \sum\limits_{s\in S_{\bar{ v}}}T_s \varepsilon_v\\
			   	\\
					C_v \gets \frac{pos + neg}{|S_d|}$\ding{182}$\\
				\END
		\END\\
	List(C_v) \gets (Normalize(List(\{C_v|\forall v\}),\lambda)\\
	\FOREACH d\in  D \DO
		\BEGIN
			norm \gets |\{s | s\in S_d, T_s \neq 0\}|\\
			\FOREACH v \in  V_d \DO
				\BEGIN
					pos \gets \sum\limits_{ s \in S_v \wedge T_s \neq 0 }(1- C_v)/T_s\\
					\\
					neg \gets \sum\limits_{ s\in S_{\bar{ v}}  \wedge T_s \neq 0}C_v/T_s\\
					\\
					\varepsilon_v \gets \frac{pos + neg}{norm}
				\END\\
			\END\\			
	List(\varepsilon_v) \gets Normalize(List(\{\varepsilon_v |\forall v\}),\lambda)\\
	\FOREACH s \in  S  \DO
			\BEGIN
				pos \gets \sum\limits_{v \in V_s \wedge \varepsilon_v \neq 0} (1-C_v)/\varepsilon_v\\
				\\
				neg \gets \sum\limits_{d\in D_s}(\sum\limits_{s\in S_{\bar{ v}}  \wedge \varepsilon_v \neq 0 } C_v/\varepsilon_v)\\
				\\
				norm \gets |\{ v \in V_{D_s} | \varepsilon_v \neq 0\}|\\
				\\
				T_s \gets  \frac{pos + neg}{norm}$\ding{183}$\\
				\\
			\END\\
List(T_s) \gets Normalize(List(\{T_s|\forall s\}),\lambda)\\
\UNTIL Convergence(T_s,\delta)\\
\FOREACH d\in D \DO
 trueValue(d) \gets  \displaystyle\argmax_{ v \in V_d} (C_v)
\end{pseudocode}
\\
\end{tabular}\vspace{-.3cm}
}
 \label{tab:algo1}
 \vspace{-.3cm}
\end{table*}

\subsection{TruthFinder}
{\scshape TruthFinder} proposed  in 2008 by Yin {\it et al.} \cite{YinHY08} applies a Bayesian analysis to compute the confidence of a claim.  

{\bf Algorithm.} 
 {\scshape TruthFinder} relies on the honesty of the sources and follows the heuristics that a source  providing mostly true claims for many data items will likely provide true claims for other objects. In Algorithm 2.1, the probability of a value being wrong is $(1 - T_s)$. Thus, if the value is provided by many sources, then its probability of being wrong is $ \prod_{s\in S_v}( 1 - T_s) $. Following this general idea, the source truthworthiness in {\scshape TruthFinder} is $T_s=\sum_{v \in V_S}C_v/|V_S|$ in \ding{183} and the confidence score of a value is $\sigma_v = -\sum_{s \in S_v}\ln (1-T_S)$. Logarithm is used to avoid underflow of the truthworthiness when the quantities are small. 
 {\scshape TruthFinder} adjusts the confidence score of a claim so that it incorporates the influence (or support) that  similar claims may have mutually on each other as $\sigma^{\star}_v=\sigma_v+\rho\sum_{v' \in V_d}\sigma_{v'}.sim(v,v')$.  For instance, for a multi-valued data item, a source providing the values {\small\texttt{(AuthorA,AuthorB)}} for a book will support another  source that provides the values {\small\texttt{(AuthorA,AuthorB,AuthorC)}} for the same book (but not inversely). The weight of such support between the values is controlled by the parameter $\rho \in [0,1]$. The final confidence of a claim is then computed in  \ding{182} with a logistic function to be positive. 
  The damping factor $ \gamma $ compensates the effect when sources with similar values are actually dependent. 
Since {\scshape TruthFinder} computes similarity between values, it can be dramatically affected by the number of distinct values to compare which explains relatively lower performance  when the number of conflicts is high. Finally, {\scshape TruthFinder} uses the difference of source truthworthiness cosine similarity between two successive iterations to be less than or equal to a given threshold, $\delta$. The value with the highest confidence is then selected as the true value among the other (false) values for a given data item.

{\bf Parameter Setting.}  
{\scshape TruthFinder} has three different parameters to be set: $\rho$, $\gamma$, and $T_S$.   
We vary every parameter value while fixing the other parameters' values as reported in the next table.

\noindent{\small
\begin{tabular}{|p{2cm}|p{1.8cm}|p{3.4cm}|p{.2cm}}
\cline{1-3}\multicolumn{1}{|c|}{{\it Fixed Values} }& \multicolumn{1}{c|}{{\it Variables}} &  \multicolumn{1}{c|}{{\it Precision}}&\\
\cline{1-3}$\rho=.5$, $T_S=.8$  & $\gamma$ from .2 to .8 &  No significant change& \\ 
\cline{1-3}$\rho=.5$, $\gamma=.1$   & $T_S$ from 0 to .99  &  No significant change&  \\ 
\cline{1-3}$\gamma=.1$, $T_S=.8$  & $\rho$ from .2 to .8  &  Max (.9777) for $\rho=.5$&\multicolumn{1}{l}{\ding{51}}  \\ 
\cline{1-3}
\end{tabular}
}\\

We vary $\delta$, the convergence threshold from $.001$ to $1$E$-5$ without any change in precision but increasing of execution time from 435~ms to 526~ms ($\approx +21$\%) for the Book data set. 
Finally, we use $\delta=.001$ and the values that maximize the precision for the Book data set: $\rho=.5$, $\gamma=.1$, and $T_S=.8$ (noted \ding{51} in the  table).
\begin{table*}
\centering
\small
\scalebox{1}{\begin{tabular}{p{5.5cm}p{4.5cm}p{8cm}}
\begin{pseudocode}{{\scshape LTM}}{S,D,V,K, burnin,thin,\alpha,\beta}
{\bf Initialization.}\\
\FOREACH d\in  D \DO
	\FOREACH v \in  V_d \DO
		\BEGIN
		C_v \gets 0\\
			\IF random( ) < 0.5 \hspace{.1cm}{\bf then   }\hspace{.1cm} t_v \gets 0 \hspace{.1cm}{\bf else     } \hspace{.1cm} t_v \gets 1\\
	\forall s \in S_v: n_{s,t_v,o_v} \gets n_{s,t_v,o_v} + 1\\
		\forall s \in S_{\bar{v}}: n_{s,t_v,o_v} \gets n_{s,t_v,o_v} + 1\\
	\END \\
\textit{\textbf{Sampling:} }\\
\FOR i \gets 1 \TO K \DO
	\BEGIN 
		i \gets i + 1\\
		\FOREACH d\in  D \DO 
			\FOREACH v \in  V_d  \DO 
				\BEGIN
 					p_{t_v} \gets \beta_{t_v} ; p_{t_{\bar{v}}} \gets \beta_{t_{\bar{v}}} \\
					\FOREACH s \in  {S_v \cup S_{\bar{v}}} \DO 
						\BEGIN
 							p_{t_v}  \gets \frac{p_{t_v}  (n_{s,t_v,o_v}+\alpha_{t_v,o_v}-1)}{n_{s,t_v,1}+n_{s,t_v,0}+\alpha_{t_v,1}+\alpha_{t_v,0}-1}\\
 							\\
 							p_{t_{\bar{v}}}  \gets  \frac{p_{t_{\bar{v}}}(n_{s,t_{\bar{v}},o_v}+\alpha_{t_{\bar{v}},o_v}-1)}{n_{s,t_{\bar{v}},1}+n_{s,t_{\bar{v}},0}+\alpha_{t_{\bar{v}},1}+\alpha_{t_{\bar{v}},0}}\\
 							\\
						\END\\
						\IF random( ) < \frac{p_{t_{\bar{v}}}}{p_{t_v}+p_{t_{\bar{v}}}} 
							\THEN 
							\BEGIN
								t_v \gets 1 - t_v \\
								\FOREACH s \in  S_v \cup S_{\bar{v}} \DO 
									\BEGIN 
								n_{s,t_{\bar{v}},o_v} \gets n_{s,t_{\bar{v}},o_v} - 1\\
										n_{s,t_v,o_v} \gets n_{s,t_v,o_v} + 1 \\
										\\
									\END \\
							\END\\
							\IF  i > burnin  \& i \% thin = 0 
								\THEN  C_v \gets C_v  + \frac{t_v.thin}{(K-burnin)}$\ding{182}$\\
				\END\\
	\END\\
	\FOREACH d\in D \\
		\hspace{.3cm}	\FOREACH v \in V_d\\
		 \hspace{.4cm}{\bf If }\hspace{.1cm}C_v > 0.5 	\hspace{.1cm}{\bf then   }\hspace{.1cm} trueValue(d)  \GETS v\\
\end{pseudocode}
&\begin{pseudocode}{{\scshape MLE}}{S,D,V,\beta_1,r,\delta}
{\bf Initialization.} \\
\FOREACH s \in S \DO
	\BEGIN
		f\gets\frac{|V_s|}{|V|}\\
		a(s) \gets r  f/\beta_1\\
		b(s) \gets (1-r)f/(1- \beta_1)
	\END\\
\REPEAT
	C_{sum} \gets 0\\
	\FOREACH  d\in  D \\
		\textit{\textbf{Expectation step: }}\\
		\FOREACH v \in  V_d \DO
			\BEGIN
				a_v \gets 1; b_v \gets 1\\
				\FOREACH s \in S_v \DO
					\BEGIN
						 a_v \gets a_v . a(s) \\
						b_v \gets b_v. b(s) \\
					\END\\
				\FOREACH s \in S_{\bar{v}} \DO
					\BEGIN
		 				a_v \gets a_v  (1 -  a(s)) \\
						b_v \gets b_v  (1 - b(s))
					\END\\
				C_v \gets \frac{a_v \beta_1}{a_v \beta_1 + b_v(1 - \beta_1)}$\ding{182}$ \\
				C_{sum} \gets C_{sum} + C_v
			\END\\
	\textit{\textbf{Maximization step: }}\\
	\FOREACH s \in S \DO
		\BEGIN
			C_{s_{sum}} \gets \sum\limits_{v \in V_s} C_v\\
			\\
			a(s) \gets C_{s_{sum}}/C_{sum}$\ding{183}$\\
			b(s) \gets |V_s| - C_{s_{sum}}/(|V| - C_{sum}) \\
		\END\\
\UNTIL Convergence(a(s),b(s),\delta)\\
\FOREACH d \in D \\
		\hspace{.3cm}	\FOREACH v \in V_d\\
	  \hspace{.4cm}{\bf If }\hspace{.1cm}C_v > 0.5 	\hspace{.1cm}{\bf then   }\hspace{.1cm} trueValue(d)  \GETS v\\
\end{pseudocode}
&\begin{pseudocode}{{\scshape Depen}}{S,D,V,n,c,\alpha,\delta}
{\bf Initialization.} \\
\forall  s \in  S:  T_s \gets 0.8 \\
\forall  d \in  D: trueValue(d)  \gets \argmax_{v \in V_d}(|S_v|)\\
\forall s_i \in S,	\forall s_j \in S-\{s_i\}:\\ \hspace{1cm}	CompDepen(s_i,s_j,\alpha,n)\\
\REPEAT
	\FOREACH d\in  D \DO
		\FOREACH  v \in  V_d \DO
		\BEGIN
 			O_{S_v} \gets orderByDepen(S_v)\\
 			Pre \gets \emptyset; 			C_v \gets 0;  t_{score_s} \gets 1; \\
 			\FOREACH s\in O_{S_v} \DO
 			\BEGIN
 				\IF Pre==\emptyset \THEN voteCount = 1 \ELSE \\
 			\hspace{.2cm}voteCount = \prod \limits_{s_j \in Pre}( 1 - ( c . depen(s,s_j)))\\
 				addToList(Pre,s)\\
 				C_v\gets C_v +  t_{score_s} . voteCount$\ding{182}$
 				\\
 			\END
		\END\\
	\FOREACH  s \in  S_v \DO
		 T_s  \gets \frac{1}{|V_s|}\sum\limits_{v \in V_s} \frac{e^{C_v}}{\sum\limits_{v' \in V_{D_v}}e^{C_{v'}}}$\ding{183}$\\
		 \\
	\forall s_i \in S, \forall s_j \in S-\{s_i\}:	CompDepen(s_i,s_j,\alpha,n)\\
	\\
\UNTIL Convergence(T_s,\delta)\\
\FOREACH d\in D\\
 trueValue(d) \gets  \displaystyle\argmax_{v \in V_d} (C_v)
\end{pseudocode}\\
 \end{tabular}
 }
 \label{tab:algo2}
 \vspace{-.3cm}
\end{table*}

\subsection{Information Corroboration}
Three algorithms have been proposed in 2010 by Galland {\it et al.} in  \cite{GallandAMS10}, namely {\scshape  Cosine}, {\scshape 2-Estimates}, and {\scshape 3-Estimates}.

\noindent{\bf {\scshape Cosine}} in Algorithm 2.2 starts by initializing the confidence of each value and the truthworthiness of each source. 
  Then,  it iteratively computes source truthworthiness in \ding{183} as a linear function of the truthworthiness achieved in the previous iteration. For each claimed value, the value confidence is computed as a function of the current truthworthiness scores of the sources claiming this value minus the truthworthiness scores of disagreeing sources in  \ding{182}.\\
\noindent{\bf{\scshape 2-Estimates}} in Algorithm 2.3 is a probabilistic model for estimating source truthworthiness and value confidence. As in {\scshape Cosine},  {\scshape 2-Estimates} takes into consideration disagreeing sources for every data item  while computing the value confidence. It starts by initializing source truthworthiness and iteratively computes the value confidence  in \ding{182} as a function of both agreeing and disagreeing sources claiming different values. Then, it computes the source truthworthiness  in \ding{183} as a function of the confidence of all values for all data items provided by the source. Finally, both value confidence and source truthworthiness are normalized after each iteration with {\it Normalize} function.\\ 
\noindent{\bf \scshape 3-Estimates} in Algorithm 2.4 uses a third parameter beside $T_S$ and $C_v$: the value error factor, $\epsilon_v$.  
Then, for each value, the algorithm computes the value confidence  in \ding{182} as a function of the value error factor and the truthworthiness of each source providing the value, as well as the truthworthiness of the sources claiming other values. The confidence is normalized and used to compute a new error factor per value which is also normalized. Finally, the source truthworthiness is computed  in \ding{183} and normalized as a function of the value confidence and the error factor. As mentioned by the authors, 
 the normalization function is critical for these algorithms to reach convergence to a non-local optima but the setting of $\lambda$ is not documented in the original paper. 
Moreover, the authors used a fix point computation for testing convergence. Since they did not recommend it for being costly and not guaranteeing the convergence in some cases, we used the same convergence test as {\scshape TruthFinder} with $\delta=.001$.

{\bf Parameter Setting.} 
 Information corroboration algorithms include four parameters to be initialized: $T_s$,  $\eta$ for  {\scshape Cosine}, $ \lambda $,  and  $\epsilon_v$. We initialize $T_s=.8$. For {\scshape Cosine}, we set $\eta$ to 0.2 since it maximizes the precision. In our parameterization analysis on the Book data set, we faced  unstable  results for {\scshape 3-Estimates}  from one execution to another 
giving different results for precision, accuracy, and recall  for certain values of $\lambda$. As shown in the table, based on 100 runs with  $\lambda=.8$, the 95\% confidence interval of precision varies from .9214 to .9587.

\noindent{\scriptsize
\begin{tabular}{|p{1.5cm}|p{1.6cm}|p{3cm}|p{.2cm}|p{.1cm}}
\cline{1-4} \multicolumn{1}{|c|}{{\it Fixed Values}   }               & \multicolumn{1}{c|}{{\it Variables} }&  \multicolumn{1}{c|}{{\it Precision}}&  \multicolumn{1}{c|}{Stability}&\\
\cline{1-4}  $\lambda=.1$, $\epsilon_v=.1$   & $T_S$ from 0 to .99  &  Max (.9805) for $T_S=.8$    & stable &  \\ 
\cline{1-4}   &    &  Min (.6647) for $\lambda=.7$   & stable &  \\ 
\cline{3-4}  $T_s=.8$, $\epsilon_v=.1$ & $\lambda$ from .1 to 1  &  Max (.9935) for  $\lambda=.5$   & stable &  \\ 
\cline{3-4}     &   &  in [.9214 to .9587] for $\lambda=.8$ & \textit{unstable} &\\ 
\cline{1-4} $T_s=.8$, $\lambda=.5$     & $\epsilon_v$ from .1 to .9  &  Max (.9935) for $\epsilon_v=.4$    & stable &\multicolumn{1}{l}{\ding{51}}\\ 
\cline{1-4}
\end{tabular}} \\

 Finally, for the Book data set, we select $T_S=.8$, $\eta=.2$ for {\scshape Cosine}, $\lambda=.5$,  and  $\epsilon_v=.4$ for {\scshape 3-Estimates}.

\subsection{Latent Truth Model}
Latent Truth Model ({\scshape LTM}) proposed in 2012 by Zhao {\it et al.}~\cite{ZhaoRGH12}  uses Bayesian networks for estimating the truth.  
{\scshape LTM} has two important assumptions on the format of the data sets it can handle: (1) the data set should contain only one attribute with atomic values and (2)  {\scshape LTM} can handle multiple true values for the same data item. For example, in the case of the Book data set where a list of authors  provided by a source $s$ is {\scriptsize\texttt{(AuthorA,AuthorB)}}, {\scshape LTM} actually takes as input two claims from  $s$, each one having an atomic value that can be true such as: {\scriptsize\texttt{(c1,s,ThisBook:AuthorOf,AuthorA)}} and  {\scriptsize\texttt{(c2,s,ThisBook:AuthorOf,AuthorB)}}. 
{\scshape LTM} considers, for each source,  its prior probability of true positive and negative errors,   noted $(\alpha_{1,1}$, $\alpha_{1,0})$  as source sensitivity, as well as its prior probability of false positive and negative errors, noted $(\alpha_{0,1},\alpha_{0,0})$ as source specificity.  Finally, values with confidence  higher than .5 are considered to be true, thus, for some data item,  LTM may not detect any true value.
%

{\bf Algorithm.} In Algorithm 2.5,  {\scshape LTM } maintains four counters for each source, noted $n_{s,t_v,o_v}$, where $t_v$ is the Boolean truth label for each value $v$, and $o_v$ is whether value $v$ is actually claimed by the source or not. {\scshape LTM } first initializes the label of each claim  randomly and updates the counters of each source. In each iteration, {\scshape LTM} samples each truth label  from its distribution conditioned on all other truth labels, and the source counters are updated accordingly. 
LTM uses a collapsed Gibbs sampling process with $K$, the  number of iterations required to define the sample size as $(K-burnin)/thin$. Then, LTM updates the values truth probability  in  \ding{182} by discarding the first set of samples ({\it burnin} parameter) and, for every $n$ samples in the remainder ({\it thin}),  {\scshape LTM} computes the average to prevent correlation between adjacent samples. 
Since {\scshape LTM} relies on the random initialization of the truth labels, as well as random sampling, we can not report the precision of one single run. In the original paper, average precision over 10 runs was reported. In our experiment, we reported the average precision over 100 runs because we observed fluctuating results with wide standard deviations over 10 runs. {\scshape LTM } does not compute source truthworthiness which gives an advantage in terms of execution time.

{\bf Parameter Setting.} Nine parameters have to be set in {\scshape LTM}: ($K$, {\it burnin}, {\it thin}): the collapsed Gibbs sampling process parameters, $\alpha=(\alpha_{1,1}$, $\alpha_{1,0}$, $\alpha_{0,1},\alpha_{0,0})$, the prior true/false positive/negative claim counts for the sources,  and $\beta=(\beta_1, \beta_0)$, the prior true and false counts for the data item-value pairs. We study the values proposed by the authors for all parameters on the Book data set: varying one parameter and fixing the others successively and we observe: (1) No significant changes in the precision of LTM, neither for ($K$, {\it burnin}, {\it thin}) = (50, 10, 1), (500, 100, 9) or  (2000, 100, 9) nor for $\beta =  (.1, .1)$ or $(.5, .5)$. (2) For high $\alpha_{0,1}$ and $\alpha_{1,0}$ (.7 to .9) and low $\alpha_{0,0}$ and $\alpha_{1,1}$ (.1 to .3),  the precision algorithm was low with  high standard deviation ($\pm 0.32$ in average) and minimal precision in the 95\% confidence interval over  100 runs. We did not consider this parameter setting for $\alpha$ because of too high variability of precision. (3) For the remaining permutations of $\alpha_{1,1}$, $\alpha_{1,0}$, $\alpha_{0,1}$, and $\alpha_{0,0}$, LTM reaches stability in  precision for 100 runs with small 95\% confidence intervals ($.002$  in average) as follows.

\noindent{\scriptsize
\begin{tabular}{|l|l|l|p{2.5cm}|p{.2cm}}
\cline{1-4}  
\multicolumn{1}{|c|}{{\it ($K, burnin, thin$)}}& \multicolumn{1}{c|}{{\it ($\beta_1$, $\beta_0$) }}    &  \multicolumn{1}{c|}{{\it ($\alpha_{1,1}$,  $\alpha_{1,0}$, $\alpha_{0,1}$, $\alpha_{0,0}$)}}&\multicolumn{1}{c|}{{\it Precision (in 95\% CI)}} &\\
\cline{1-4}
    &  (.1, .1)& \multirow{2}{*}{(.9, .1, .9, .1)}& Max [.8556;.8580]&\\
    \cline{2-2}    
\cline{2-2}        (50, 10, 1) &  (.5, .5)&  & Max [.8563;.8585]& \\
 \cline{2-2}\cline{3-3}  \cline{4-4}    &  (.1, .1)& \multirow{2}{*}{(.1, .9, .9, .1)}& Min  [.6851;.7953]&\\
\cline{2-2}  
\cline{2-2}  \cline{4-4}         &  (.5, .5)& & Min  [.6636;.7812]&\\
\cline{1-4}
\cline{2-2} \cline{4-4}      &  (.1, .1)& \multirow{2}{*}{(.9, .1, .9, .1)}& Max [.8588;.8610]&\ding{51}  \\
\cline{2-2} \cline{4-4}      (500, 100, 9) &  (.5, .5)& &  [.8579;.8601] & \\
\cline{2-2}\cline{3-3}  \cline{4-4}    &  (.1, .1)& \multirow{2}{*}{(.1, .9, .1, .9)}&Min [.8515;.8539]&\\
\cline{2-2} \cline{4-4}    &  (.5, .5)& &  [.8521;.8534]&\\
\cline{1-4}
\end{tabular}
}\\

Finally, we select ($K$, {\it burnin}, {\it thin}) = (500, 100, 9), $\alpha=(.9, .1, .9, .1)$ and $\beta=(.1, .1)$ to get maximal precision average over 100 runs on the Book data set.

\begin{table*}
\centering
\small
\scalebox{0.9}{
\begin{tabular}{p{5.8cm}p{5.2cm}p{1.4cm}|p{1cm}|p{1.35cm}|p{.65cm}}
\cline{4-6}
\multirow{11}{*}{\begin{pseudocode}{SimpleLCA}{S,D,V,W,\beta_1,\delta}
 \forall s \in S: T_S \gets .8\\
\REPEAT
\FOREACH d\in  D \\
\textit{\textbf{Expectation step: }}\\
\BEGIN
C_{d_{sum}} \gets 0\\
\FOREACH v \in  V_d \DO
\BEGIN
C_v \gets \beta_1 . \prod\limits_{s \in S_v} T_s^{w_{s,d}}  \\
 \hspace{.7cm} . \prod\limits_{s' \in S_{\bar{v}}} ((1 - T_{s'})/(|V_d| - 1))^{w_{s,d}} $\ding{182}$\\
\\
C_{d_{sum}}  \gets C_{d_{sum}}  + C_v\\
\END\\
\FOREACH v \in  V_d \DO
C_v \gets C_v/C_{d_{sum}}$\ding{182}$
\END\\
\textit{\textbf{Maximixation step: }}\\
\FOREACH s \in  S_v \DO T_s \gets \sum\limits_{v \in V_s} C_v.w_{s,d}/ \sum\limits_{d \in D} w_{s,d}$\ding{183}$ \\
\UNTIL Convergence(T_s,\delta)\\
\FOREACH d\in D:  trueValue(d) \gets  \displaystyle\argmax_{v \in V_d} (C_v)\\
\end{pseudocode}}
&\multirow{11}{*}{\begin{pseudocode}{GuessLCA}{S,D,V,W,\beta_1,\delta}
 \forall v: p_{g_v} \gets |S_v|/(|S_v| + |S_{\bar{v}}|) \\
  \forall s \in S: T_S \gets .8\\
\REPEAT 
\FOREACH  d\in  D \\
\textit{\textbf{Expectation step: }}\\
\BEGIN
 C_{d_{sum}}  \gets 0 \\
\FOREACH v \in  V_d \DO
\BEGIN
 C_v \gets \beta_1 .   \prod\limits_{s \in S_v} (T_s +  (1 - T_s)p_{g_v} )^{w_{s,d}} \\
 \hspace{.7cm} .  \prod\limits_{s' \in S_{\bar{v}}} ( (1 - T_{s'}) p_{g_v})^{w_{s,d}} $\ding{182}$ \\
 \\
 C_{d_{sum}}  \gets C_{d_{sum}}  + C_v \\
 \\
\END\\
\FOREACH v \in  V_d:  C_v \gets C_v/C_{d_{sum}}  
\END\\
\textit{\textbf{Maximixation step: }}\\
\FOREACH s \in  S_v \DO 
 T_s \gets (\sum\limits_{v \in V_s} C_v + \sum\limits_{v \in V_{D_s}} \frac{p_{g_v}}{1-p_{g_v}} C_{v})/\big(\sum\limits_{v \in V_{D_s}} C_{v}.w_{s,d}\big)\big)$\ding{183}$ \\
\UNTIL Convergence(T_s,\delta)\\
\FOREACH  d \in D:  trueValue(d)  \gets   \displaystyle\argmax_{v \in V_d} (C_v) \\
\end{pseudocode} } 
& 	&	\multicolumn{1}{c|}{{\bf	  Confidence }} & \multicolumn{1}{c|}{ {\bf Truthworthiness }} & \multicolumn{1}{c|}{{\bf Time }}\\
& &	&	\multicolumn{1}{c|}{{\bf	   Computation  }}& \multicolumn{1}{c|}{ {\bf  Computation }} & \multicolumn{1}{c|}{{\bf  Complexity}}\\
& &	&\multicolumn{1}{c|}{\ding{182}	}&\multicolumn{1}{c|}{\ding{183}} & \multicolumn{1}{c|}{{\bf  per Iteration}}\\
\cline{3-6}
&&\multicolumn{1}{|c|}{}&&&\multicolumn{1}{c|}{}\\
&&\multicolumn{1}{|l|}{{\scshape  Voting    }}                & \multicolumn{1}{c|}{$|  S_v|.|V|$    }    &    \multicolumn{1}{c|}{    -}                &    \multicolumn{1}{c|}{$|S_v|.|V| $}\\
&&\multicolumn{1}{|l|}{{\scshape TruthFinder}     }        & \multicolumn{1}{c|}{$| S_v|.|V|$    }    & \multicolumn{1}{c|}{$ |S|.|V_s| $}    &    \multicolumn{1}{c|}{    $ |S|.|V|+|V_d|^2 $}\\
&&\multicolumn{1}{|l|}{{\scshape Cosine}     }                    & \multicolumn{1}{c|}{$ |S|.|V|+ |V| $} & \multicolumn{1}{c|}{$ |S|.|V| + |S| $}&    \multicolumn{1}{c|}{$ |S|.|V| $}  \\
&&\multicolumn{1}{|l|}{{\scshape 2-Estimates }}            & \multicolumn{1}{c|}{$ |S|.|V| + |V| $}& \multicolumn{1}{c|}{$ |S|.|V| + |S| $}& \multicolumn{1}{c|}{$ |S|.|V| $} \\
&&\multicolumn{1}{|l|}{{\scshape 3-Estimates}}             & \multicolumn{1}{c|}{$ |S|.|V| + |V| $}& \multicolumn{1}{c|}{$ |S|.|V| + |S| $}&\multicolumn{1}{c|}{ $ |S|.|V| $}\\
&&\multicolumn{1}{|l|}{{\scshape LTM }}                            & \multicolumn{1}{c|}{$ |S.|V| $ }        &    \multicolumn{1}{c|}{- }                  & \multicolumn{1}{c|}{$ |S|.|V| $}\\
&&\multicolumn{1}{|l|}{{\scshape MLE }}                         & \multicolumn{1}{c|}{$ |S|.|V|  $}        & \multicolumn{1}{c|}{$ |S|.|V_s| $}    &\multicolumn{1}{c|}{$ |S|.|V| $}\\
&&\multicolumn{1}{|l|}{{\scshape Depen}}                         & \multicolumn{1}{c|}{$ |S_{v}|^2.|V| $}&\multicolumn{1}{c|}{ $|S|.|V| + |S|^2.|V_{s}|^2 $}&\multicolumn{1}{c|}{$ |S|^2.|V_s|^2$ }\\
&&\multicolumn{1}{|l|}{{\scshape Accu}}                         & \multicolumn{1}{c|}{$ |S_{v}|^2.|V| $}&\multicolumn{1}{c|}{ $|S|.|V| + |S|^2.|V_{s}|^2 $}&\multicolumn{1}{c|}{$ |S|^2.|V_s|^2$ }\\
&&\multicolumn{1}{|l|}{{\scshape AccuSim}}                     & \multicolumn{1}{c|}{$ |S_{v}|^2.|V| $}&\multicolumn{1}{c|}{ $|S|.|V| + |S|^2.|V_{s}|^2 $}&\multicolumn{1}{c|}{$ |S|^2.|V_s|^2 + |V_d|^2$ }\\
&&\multicolumn{1}{|l|}{{\scshape SimpleLCA} }    &\multicolumn{1}{c|}{$  |S|.|V| $}     &  \multicolumn{1}{c|}{ $ |S|.|V_s| $}                 &\multicolumn{1}{c|}{ $ |S|.|V| $}\\
&&\multicolumn{1}{|l|}{{\scshape GuessLCA}}    & \multicolumn{1}{c|}{$ |S|.|V|$ }        & \multicolumn{1}{c|}{$| S|.|V| $}                         &\multicolumn{1}{c|}{$ |S|.|V| $}  \\
\cline{3-6}
\multicolumn{6}{l}{}\\
&&\multicolumn{4}{l}{\hspace{2cm}{\normalsize{\bf Table 3. Time Complexity Analysis  }}}\\
\multicolumn{6}{l}{}\\
\multicolumn{6}{l}{}\\
\multicolumn{6}{l}{}\\
\multicolumn{6}{l}{}\\
\multicolumn{6}{l}{}\\
\end{tabular}
}
\end{table*}
%
\subsection{ Maximum Likelihood Estimation}
Maximum Likelihood Estimation ({\scshape MLE}) proposed in 2012 by Wang {\it et al.} in \cite{WangKLA12}  is based on the Expectation Maximization (EM) algorithm to quantify the reliability of sources and the correctness of their observations. {\scshape MLE} only deals with Boolean positive observations (e.g., data items such as {\scriptsize\texttt{thisPerson-hasKids}} with  {\scriptsize\texttt{True}} or {\scriptsize\texttt{False}} value). Negative observations are ignored. 
To be able to test {\scshape MLE} on the Book data set, we reformated every claim such as {\scriptsize\texttt{(c1,s,ThisBook:AuthorOf,(AuthorA,AuthorB))}} such as two claims:  {\scriptsize\texttt{(c1,s,ThisBook:AuthorOf:AuthorA,True)}} and {\scriptsize\texttt{(c2,s,ThisBook:AuthorOf:AuthorB,True)}}. 

{\bf Algorithm.} In Algorithm 2.6,  {\scshape MLE} starts with initializing the sources' parameters: $a(s)$, the probability that  source $s$ reports a value to be true when its indeed true and $b(s)$, the probability that $s$ reports a value to be true when it is in reality false (similar to source sensitivity $\alpha_{1,1}$ and $\alpha_{1,0}$ in {\scshape LTM}). In the Expectation step, {\scshape MLE} iteratively computes the conditional probability of a value $v$ to be true based on its source probabilities ($a(s)$, $b(s)$), and on the probabilities of the sources not providing $v$ ($\forall s \in S_{\bar{v}}$).  Then, it iteratively computes the confidence of each value in~\ding{182}. 
 In the Maximization step, {\scshape MLE} updates the sources probabilities $a(s)$ and $b(s)$ in~\ding{183}.  
 The Expectation-Maximization steps are repeated until convergence of both $a(s)$ and $b(s)$. 
  An important observation of MLE algorithm is when the number of sources tends to be very large, source probabilities tend to zero and $C_v$  tends to $0/0$. MLE can not be used with a large number of sources ($>$ 5,000).

{\bf Parameter Setting.} Two parameters are needed in MLE: $r$ and $\beta_1$  to compute the initial parameters of the sources, $a(s)$ and $b(s)$. $\beta_1$ is the overall prior truth probability of the claims (similarly to {\scshape LTM}). $r$ is the probability that a source provides a value for all data items. In its original paper,  {\scshape MLE} is tested  on a synthetic data set with no indication on how to set these parameters. So, for the Book data set, we successively vary $r$ and $\beta_1$  using a uniform constant value  for all sources parameters initialization.

\noindent{\scriptsize
\begin{tabular}{|p{2.3cm}|p{1.6cm}|p{3.2cm}|p{.2cm}}
\cline{1-3} \multicolumn{1}{|c|}{{\it Fixed Values} }         & \multicolumn{1}{c|}{{\it  Variables}} &\multicolumn{1}{c|}{{\it  Precision, Accuracy, Recall} }& \\
\cline{1-3}$r=.5$ for all sources&   $\beta_1$ from .1 to .9 &  All equal to 1 for $\beta_1=.5$&\ding{51} \\
 \cline{1-3}
       $\beta_1=.5$ & $r$ from .1 to .9 & All equal to 1 for $r=.5$& \\
\cline{1-3}
\end{tabular}}\\
Finally, we select $\beta_1=.5$ and  $r=.5$ uniformly constant for all sources to get  precision, accuracy, and recall equal to 1.
\subsection{Source Dependence in Truth Discovery}
{\scshape Depen} proposed in 2009 by Dong {\it et al.}  \cite{DongBS09} and further extended in \cite{DongBS09a,Dong:2010} is the first Bayesian truth detection model that takes into consideration the copying relationships between sources. {\scshape Depen} penalizes the vote count of a source if the source is detected to be a copier of another source. 
  {\scshape Depen} is presented with 4 extensions in its original paper \cite{DongBS09}. 
Our study focuses on {\scshape Depen}, {\scshape Accu}, {\scshape AccuSim}, and  {\scshape AccuNoDep}: {\scshape Accu} extends {\scshape Depen} model and relaxes the assumption that the sources have the same accuracy and for each data item, all independent sources have no longer the same probability of providing a true value. {\scshape AccuSim} extends {\scshape Accu} to take into account value similarity, and  {\scshape AccuNoDep} assumes that all sources are independent.

{\bf Algorithm.} Algorithm 2.7 presents  {\scshape Depen} and starts by initializing all sources' truthworthiness to .8. For every data item, it selects the true value by majority voting, and computes the dependence between sources with  $CompDepen(s_i,s_j,\alpha,n)$ function  where 
 $\alpha$ is  the {\it a priori} probability that $s_i$ and $s_j$ are dependent, and $n$ is the number of false values per data item. To iteratively compute the value confidence   in~\ding{182}, the sources claiming the considered value are first ordered by their dependence probabilities with {\it orderByDepen(S$_v$)} function. Then, each source's {\it voteCount} is computed in a way that minimizes the vote if the source is dependent on other sources in {\it Pre}, the  list of  ranked sources, such as {\it voteCount}$=\prod_{s_j\in Pre}(1-c.${\it depen(s,s$_j$))}, with $c$ the probability that a value provided by a copier is copied. {\it voteCount} is then weighted by $t_{score_s}$, the source's  score  to compute the value confidence.   Source truthworthiness is computed iteratively in~\ding{183} as a function of the confidence of  all values claimed by the sources. True values are expected to be the values with the highest confidence. 
In {\scshape AccuNoDep}, no dependence computation is needed, and {\it voteCount} is always 1. 
In {\scshape Accu} and {\scshape AccuSim}, the algorithm computes value confidence  with $t_{score_s}= \ln (n T_s/(1 - T_s)) $, whereas in {\scshape Depen}, $t_{score_s} = 1$. 
In {\scshape AccuSim},  the value similarity is considered for the confidence computation in each iteration and,  $ \rho \sum_{v'\in V_d}C_{v'}. sim(v',v)$ is added to $C_v$ (similarly to {\scshape TruthFinder}).  It is worth noticing that {\scshape Depen} model and its extensions estimate the source {\it voteCount} for a given value based on ordering the sources by decreasing dependence probability. This ordering could be different from one run to the next, because two sources with the same dependence probabilities could appear in different positions. We observed that this dependence-based ordering introduced small fluctuations of  the quality metrics for 20 executions of the models with the same parameterization on the Book data set. In particular, we observe  {\scshape Depen} precision ($.9814\pm.0002$), {\scshape Accu} precision ($.9741\pm.0061$) and {\scshape AccuSim} precision ($.9413\pm.0051$).  To mitigate this problem, we decided to use the lexical ordering rather than the dependence probability-based ordering of the sources. This sightly improves  the quality  of the models by +.02 ({\scshape Depen$_{LEX}$} precision $.9814$, {\scshape Accu$_{LEX}$}  precision $.9809$, and {\scshape AccuSim$_{LEX}$}  precision 0.973) for the Book data set and it also improves the stability of the results that remain constant from one run to another. 

{\bf Parameter Setting.}  
Fixing $\rho=.5$ and $n = 100$, we study various parametrization setting reported in the table.

\noindent{\scriptsize \begin{tabular}{|l|l|p{4.1cm}|l}
\cline{1-3}\multicolumn{1}{|c|}{{\it Fixed Values}} & \multicolumn{1}{c|}{ {\it Variables}} & \multicolumn{1}{c|}{{\it Precision}}&\\
\cline{1-3}
 $\alpha=.2$, $c=.8$&$T_s$ from 0 to .99&   {\scshape Depen}: Max (.9814) for  $T_s=.8$   &         \\
\cline{1-3}$T_s=.8$, $c=.8$         & $\alpha$ from .1 to.5  & {\scshape Depen}: Max (.9814) for $\alpha=.2$      &         \\
\cline{1-3}  $T_s=.8$, $\alpha=.2$   &        $c$ from .05 to .95       &  {\scshape Depen} \& {\scshape AccuNoDep}: Max (.9814)       & \ding{51}  \\
                                             &                                   & for $c=.8$;  {\scshape Accu}: Max (.9809) for $c=.1$ &   \\
     && {\scshape AccuSim}: Max (.973) for $c=.05$       &   \\
\cline{1-3}
\end{tabular}}\\

Finally, we select $\alpha=.2$, $ T_S =.8$,  $ c=.8$ for  {\scshape Depen} and {\scshape AccuNoDep} and $c=.1$ for {\scshape Accu} and $c=.05$ for {\scshape AccuSim}. 

\newpage
\subsection{Latent Credibility Analysis }
Latent Credibility Analysis ({\scshape LCA}) proposed in  2013 by Pasternack and Roth in \cite{PasternackR13} is a probabilistic model that also uses the Expectation Maximization algorithm to calculate the probability of a claim being true, by grouping claims related to the same data items into mutual exclusion sets where only one true claim exists. 
 Four LCA variants have been proposed in the original paper. 
  In our study, we focus on: {\scshape SimpleLCA} and {\scshape GuessLCA}. Both algorithms require $W$, a 
    confidence matrix that expresses the confidence of each source $s$ in its assertions for each data item $d$ (with {$w_{s,d}$ elements). Typically, $w_{s,d}$  will be 1 if the source $s$ asserts with full certainty a value for $d$, or 0 if the source says nothing about $d$. \\
\noindent{\bf SimpleLCA} is the simplest and straightforward approach where each source has a probability of being honest and all sources are considered to be independent.  In the Expectation step of Algorithm 2.8,  {\scshape SimpleLCA} iteratively computes the confidence of each value in ~\ding{182}  with $\beta_1$, the prior truth probability of the claimed value (similarly to LTM and MLE). 
  Then,  {\scshape SimpleLCA} iteratively computes the source truthworthiness in the Maximization step in~\ding{183}, in the same way as {\scshape TruthFinder},  averaging the confidence of the values that the source provides weighted by the certainty of the source on each of its assertions.\\
\noindent{\bf GuessLCA}.  {\scshape GuessLCA} extends {\scshape SimpleLCA} with the probability of a source guessing when being honest, noted $p_{g_v}$. {\scshape GuessLCA} rewards hard claims with correct truth label and penalizes getting easy claims wrong. It also assumes that no source will do worse than guessing, which is a significant advantage over other methods for pessimistic scenarios, as we will see in the next section. $p_{g_v}$ can be uniformly constant or set to the distribution of sources asserting the claims for a given data item. The main  assumption is that a guessing source chooses randomly according to the distribution of votes. 
  In Algorithm 2.9, the confidence of value $v$  is computed in~\ding{182} as the product of  $\beta_1$ with the probability that the sources assert $v$ as a true claim knowing the truth and also guessing  as $T_S+(1-T_S)p_{g_v}$, and the probability of not knowing the truth and guessing as $(1-T_{S'})p_{g_v}$ for $s' \in S_{\bar{v}}$ to the power $w_{s,d}$, the source's confidence in the value it claims for each data item. Source truthworthiness is computed in~\ding{183}. Convergence test for the {\scshape LCA} models was not explicitly mentioned in the original paper, only the required number of iterations was stated to be 50 iterations. In our experiments, we use the same convergence test as for the other methods: the difference of cosine similarity of both source truthworthiness and value confidence between two iterations, to be less than or equal to $\delta=.001$. 

{\bf Parameter Setting.} Similarly to {\scshape LTM} and {\scshape MLE}, {\scshape LCA} models require,  as input parameters, the prior truth probability  $\beta_1$ and the honesty of the sources, noted $T_S$ in our notation. We tested various parameter settings on the Book data set. We finally select $\beta_1=.5$ and $T_S=.8$ for maximizing precision of {\scshape LCA} models.

\noindent{\scriptsize
\begin{tabular}{|p{1.4cm}|p{1.6cm}|p{4.1cm}|l}
\cline{1-3} \multicolumn{1}{|c|}{{\it Fixed Values}} & \multicolumn{1}{c|}{{\it Variables} }& \multicolumn{1}{c|}{{\it Precision}}&\\
\cline{1-3} $T_s=.8$     & $\beta_1$ from .1 to 1 &  {\scshape GuessLCA}: Max (.9806) and {\mbox{\scshape SimpleLCA}:} Max (.9758) for  $\beta_1=.5$& \ding{51} \\
\cline{1-3}
\end{tabular}}\\

\subsection{Conclusions on Parameter Setting}~\label{sect:conclusion1}
The main conclusions of our parameterization study are mainly related to the modeling assumptions, the usability of the algorithms, and the repeatability of the results.

{\bf (1) Modeling Assumptions.}  
First, all methods rely on various assumptions that have direct impact on the quality and applicability of the models:  ({\bf A1})-- a source is supposed to contribute uniformly to all the claims it expresses. In every algorithm,  $T_S$ and  {\it a priori} probabilities are uniformly distributed either across all sources or all claims. As a consequence, the models do not explicitly consider both the expertise of certain sources (which can be either general or more specialized on particular topics or claims) and the hardness of certain claims (except {\scshape 3-Estimates} or  {\scshape GuessLCA}). Only {\scshape LCA} models express the degree of certainty some sources may have on their claimed values.  
({\bf A2})-- Concerning the type of the claims as inputs of the algorithms: all claims are assumed to be positive and directly attributed to a source, {\it i.e.,}  cases such  as ``$S$ claims that $A$ is false'', or ``$S$ does not claim $A$ is true'', or ``According to $S_1$, $S_2$ claims that $A$ is true'' are not considered in the models we studied. For LTM, claim structure is restricted to single-property assertions and MLE requires Boolean values to be comparable with other algorithms. This requisites may cause information omissions or distortions  due to data formatting. Except for LTM, ({\bf A3})-- all models consider that exactly one of the claims for a given data item has to be true. Thus, multiple views of the truth are not modeled.  None of the models penalize the sources claiming multiple values (similar or distinct) for the same data item. Except {\scshape Depen} and its recent extensions in \cite{Pochampally2014}, ({\bf A4})-- sources and claims are assumed to be independent, as well as real-world objects they refer to.

\begin{table*}
\small
\centering\resizebox{0.82\textwidth}{!}{\begin{minipage}{\textwidth}
\begin{tabular}{|p{5cm}|p{3.8cm}|p{12.3cm}|}
\hline \multicolumn{1}{|c|}{{\bf Control Parameter}}& \multicolumn{1}{c|}{{\bf  Value }}&\multicolumn{1}{c|}{{\bf Description} }\\
\hline \texttt{Number of sources} \texttt{(S)}& 50 ; 1,000 to 10,000 & The number of sources providing claims: $|S|=50$  in Section 3.1 and from 1,000 to 10,000 in Section 3.2.\\
\hline \texttt{Number of data items} \texttt{(D)}& 1,000 ; 100 to 10,000 & The number of data items, {\it i.e.}, pairs of (object,attribute) with claimed values: $|D|=$1,000 in Section 3.1 and from 100 to 10,000 in Section 3.2. \\
\hline \texttt{Source Coverage} \texttt{(Cov)}& \texttt{U25; U75}\hspace{.05cm}  (Uniform)& The number of  values provided by the sources is uniformly distributed on 25\% and 75\% of the data items. \\
 &\texttt{E}\hspace{.4cm} (Exponential)& The number of  values provided by the sources is exponentially distributed across the data items.\\
\hline 
\texttt{Ground Truth Distribution }& {\texttt R}\hspace{.2cm}  (Random)& The number of true positive claims per source is random. \\
\texttt{ per Source (GT)}&\texttt{U25;  U75 }(Uniform)& Each source provides the same number of true positive claims.\\
& \texttt{FP} \hspace{.2cm} (Fully Pessimistic)&80\% of the sources provide always false claims and 20\% of the sources provide always true positive claims.\\
& \texttt{FO} \hspace{.2cm} (Fully Optimistic)&80\% of the sources provide always true positive claims and 20\% of the sources provide always false claims.\\
& \texttt{80P} \hspace{.2cm} (80-Pessimistic)&80\% of the sources provide 20\% true positive claims. 20\% of the sources provide 80\% true positive claims.\\
& \texttt{80O} \hspace{.2cm} (80-Optimistic)&80\% of the sources provide 80\% true positive claims. 20\% of the sources provide 20\% true positive claims.\\
&\texttt{E} \hspace{.2cm} (Exponential)& The number of true positive values provided by the sources is exponentially distributed. 
\\
\hline
\texttt{Distinct Value Distribution }& \texttt{U}\hspace{.4cm}  (Uniform) &All data items have the same number of distinct values claimed by the set of sources.\\
\texttt{ per Data item (Conf)} &\texttt{E}\hspace{.4cm}  (Exponential)& Each data item has a number of distinct values that is exponentially distributed.\\
\hline

\texttt{Number of Distinct Values}& $2 \dots 20$ & The number of distinct values per data item. \\
\hline
\end{tabular}\hfill\
      \end{minipage}
      }\hfill\
\caption*{{\bf Table 4. Parameters for Synthetic Data Sets Generation for Configurating a Truth Discovery Scenario}}\label{tab:synthetic-data set}
\end{table*}

{\bf (2) Usability.} 
Our main observation is that all models require complex, {\it ad hoc} parameter setting and tuning depending on the considered data set.  We observe that the parameter settings we selected to maximize precision for the Book data set do not maximize precision of the algorithms when they are applied to other data sets. The gold standard  of the Book data set represents 7.91\% of the data set. We  argue that it is not representative enough for a systematic, rigorous comparison of the algorithms' quality.  Optimal parameterization of the algorithms based on a real-world data set is jeopardized when the ground truth is {\it partial} and reduced to samples of the real-world data set. This problem actually motivated us to develop a framework and a  synthetic data set generator to systematically control the  {\it complete } ground truth distribution, as we will describe in the next section. 

  {\bf (3) Repeatability.}  
   We make several observations from the parameterization study on the Book data set. First, $T_S$ initialization (uniformly constant across the sources) generally does  not have an impact on the algorithms' precision. Most importantly, we observe unstable results of {\scshape 3-Estimates} and LTM over multiple runs. The authors of {\scshape 3-Estimates} introduced a  normalization function to reach convergence but the  parameter setting ($\lambda$) of this function generates results that are not reproducible from one execution to another. Due to randomization, LTM requires  more than $100$ executions to reduce the standard deviation and 95\% confidence interval of the averaged precision, but only for certain settings of $\alpha_1$ (source sensitivity) on the Book data set, regardless of the number of LTM iterations or $\beta$ settings. 
     Two other important observations concern the computational issues and convergence of  MLE, LCA, and LTM. As a common problem in Bayesian computation, the product of prior probabilities may be too small to be represented as a floating point number and the calculation involving these numbers may underflow to zero and produce NaN results.  MLE and LCA algorithms suffer from this problem when the number of sources is greater than 5,000. One way to overcome this problem is to use $\log$ in the value confidence computation (similarly to {\scshape TruthFinder}). Concerning  convergence, we use the difference of sources' truthworthiness cosine similarity between two successive iterations to be less than or equal to $\delta=.001$ as a convergence test for all algorithms. However, LTM requires a number of iterations $K$ as input. Although LTM may reach maximal precision after few runs, it iterates until $K$ and requires multiple runs ($>$ 100 runs). As we will see in the next section, convergence of LCA models is not reached in certain cases after 500 iterations, which is the fixed limit in our experiments. 
\section{Comparative Experiments}~\label{sect:exp}
%
A first set of experiments has been conducted over synthetic data sets  to evaluate the quality (Section 3.1) and scalability of each algorithm (Section 3.2). A second set of experiments has been conducted over five real-world data sets to report the running time, number of iterations, and memory usage in addition to each algorithm's quality metrics (Section 3.3).  
 Quality of the algorithms is measured with four metrics computed  either from the gold standard in the case of real-world data sets, or from the ground truth in synthetic data sets as  
  {\small
\begin{tabular}{l c l}
  {\it  Precision} $= \frac{TP}{TP+FP} $ & \qquad \qquad&  {\it Accuracy} $= \frac{TP+TN}{TP+FP+TN+FN} $ \\\\
   {\it Recall }$= \frac{TP}{TP+FN} $    &              &  {\it Specificity} $= \frac{TN}{FP+TN} $\\
\end{tabular}}\\
with \\
\noindent{\scriptsize \begin{tabular}{p{1cm} r|r|c|c|}
\cline{4-5}    & \multicolumn{2}{c|}{}        &   \multicolumn{2}{c|}{Ground Truth / Gold Standard}  \\
\cline{4-5}    & \multicolumn{2}{c|}{}        & True & False\\
\cline{2-5}&\multicolumn{1}{|c|}{Algorithm}& True & True Positive (TP) &  False Positive (FP)\\
\cline{3-5}  &\multicolumn{1}{|c|}{}&False& False Negative (FN)& True Negative (TN)\\
\cline{2-5}
\end{tabular}}\\

The execution time is the total time to compute the truth discovery results, including initialization, convergence, eventual normalization, computation of source truthworthiness and value confidence. 
 We re-implemented all the algorithms in Java 7 under a common implementation framework to test as accurately as possible their relative quality, performance,  and behavior. Source codes are available in~\cite{BW2014}.   We ran experiments on 3 PCs with Intel Core i7-2600 processor (3.40GHz$\times$8, 32GB).

\subsection{ Experiments on Synthetic Data} 

First, we generated synthetic data to evaluate the algorithms under a
wide range of truth discovery scenarios. Table~4 summarizes the parameters we used to control the characteristics of the synthetic data set generation. 
In particular, we  control the percentage and distribution of data items for which a source  claims a value {\scriptsize\texttt{(Cov)}} and the  number and distribution model of distinct values per data item {\scriptsize\texttt{(Conf)}}. 
 We also control the percentage and distribution model of true positive values per source {\scriptsize\texttt{(GT)}}. This actually constitutes the ground truth we used for computing the quality metrics  of the algorithms. 
  Finally, we ran our experiments on 9,120 data sets generated with $|S|=50$ and $|D|=1,000$: 10 data sets for each of the $(3 \times 8 \times 2 \times 19) $ possible configurations presented in  Table 4. Due to space limitation, only 8 configurations are presented in this section and in Fig. 1 -- see~\cite{BW2014} for more detailed and experimental results. Dependence between sources and value similarity were not the scope of this study since these aspects are considered only by {\scshape TruthFinder}, {\scshape AccuSim}, and {\scshape Depen}.  
  In the set of experiments on synthetic data, our objective is to identify the data set characteristics that have an impact on the quality of the algorithms, in particular: (1) the number of values claimed by the sources; (2) the number and distribution of distinct values per source, and (3) the type of distribution of true positive claims per data item. Fig. 1 shows the algorithms' precision average over 10 data sets for each configuration with an increasing number of distinct values per data item (from 2 to 20).

\begin{figure*}
\centering
\subfigure[\scriptsize{\texttt{Cov=U25\&U75-Conf=U\&E-GT=R}}]{
\includegraphics[width=.24\linewidth]{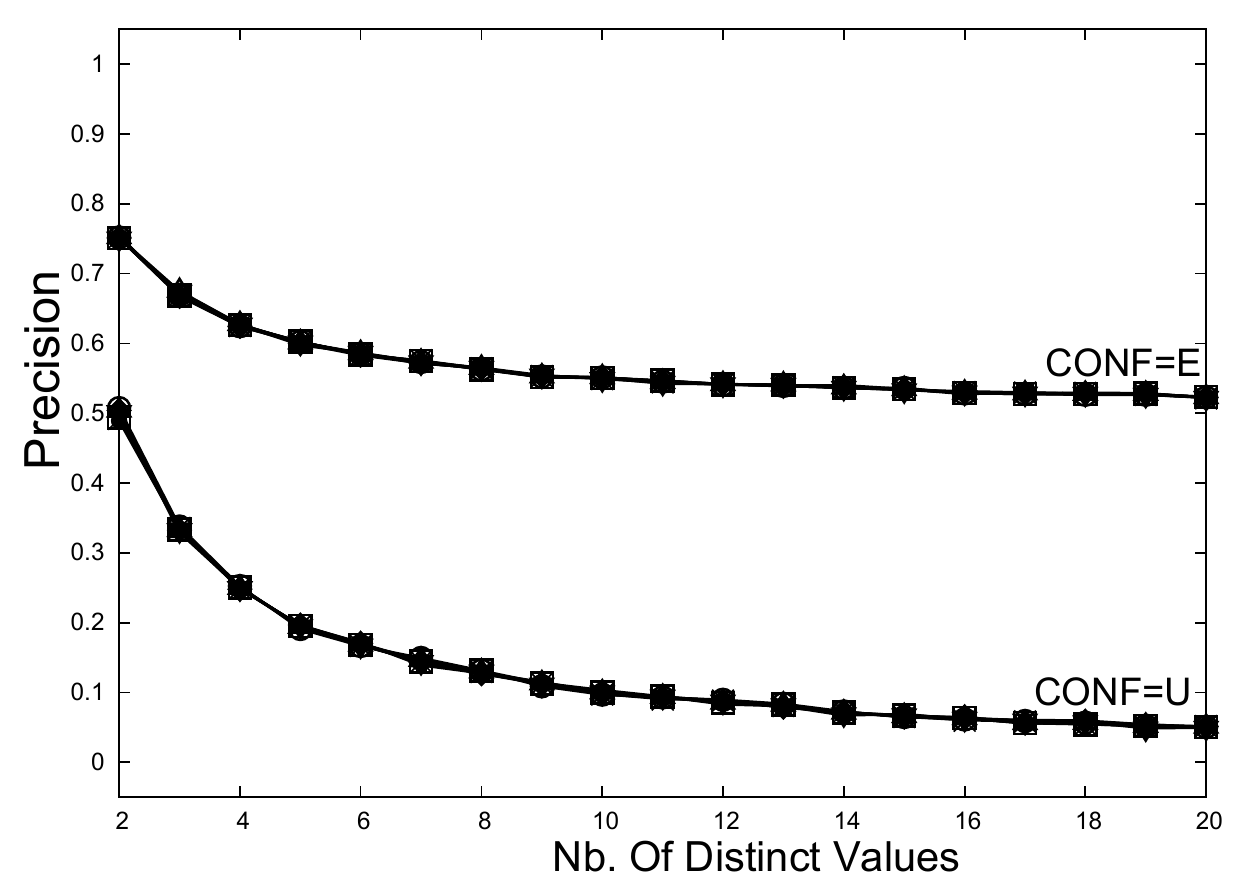}}\label{fig:U25U75-UE-GT=R}
\subfigure[\scriptsize{\texttt{Cov=E-Conf=E-GT=U25}}]{
\includegraphics[width=.24\linewidth]{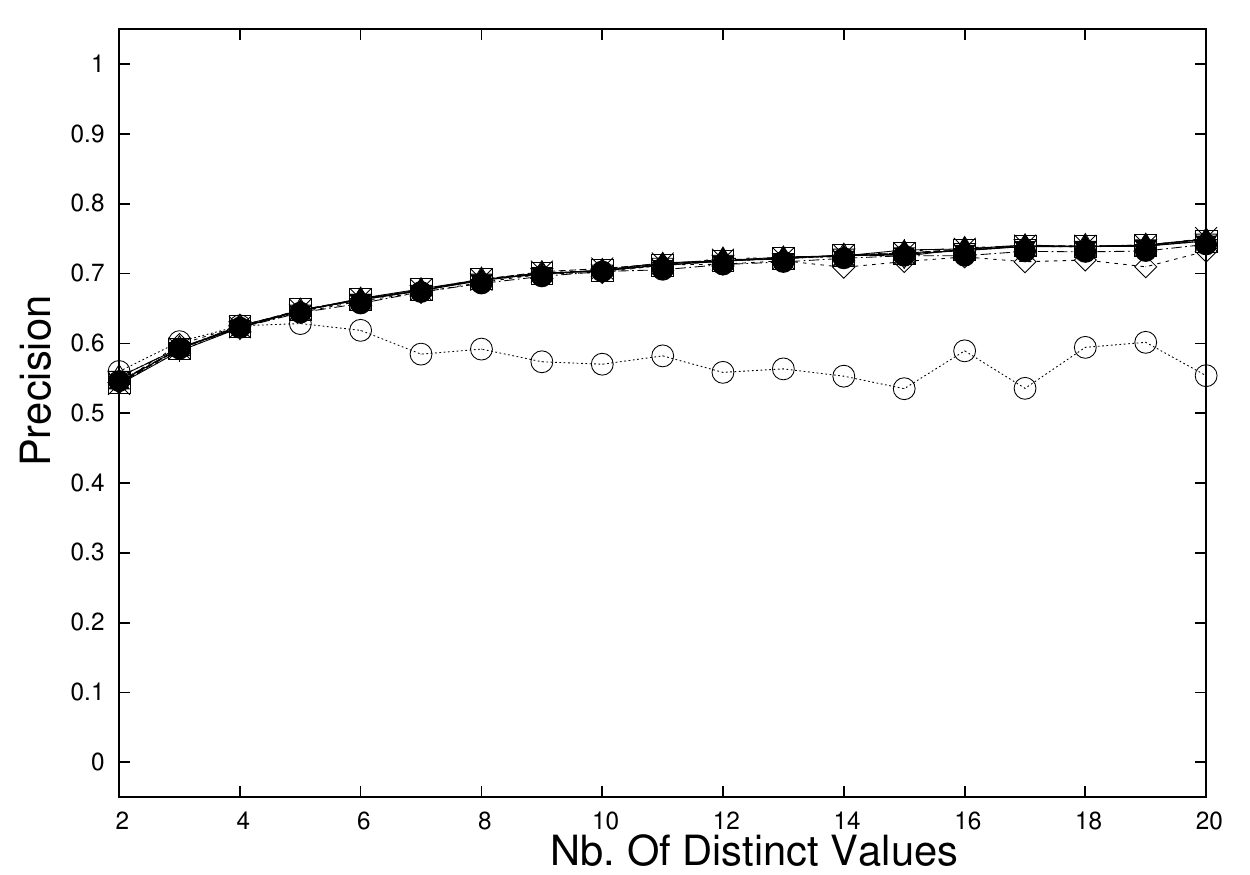}}\label{fig:EEU25}
\subfigure[\scriptsize{\texttt{Cov=E-Conf=U-GT=FP}}]{
\includegraphics[width=.24\linewidth]{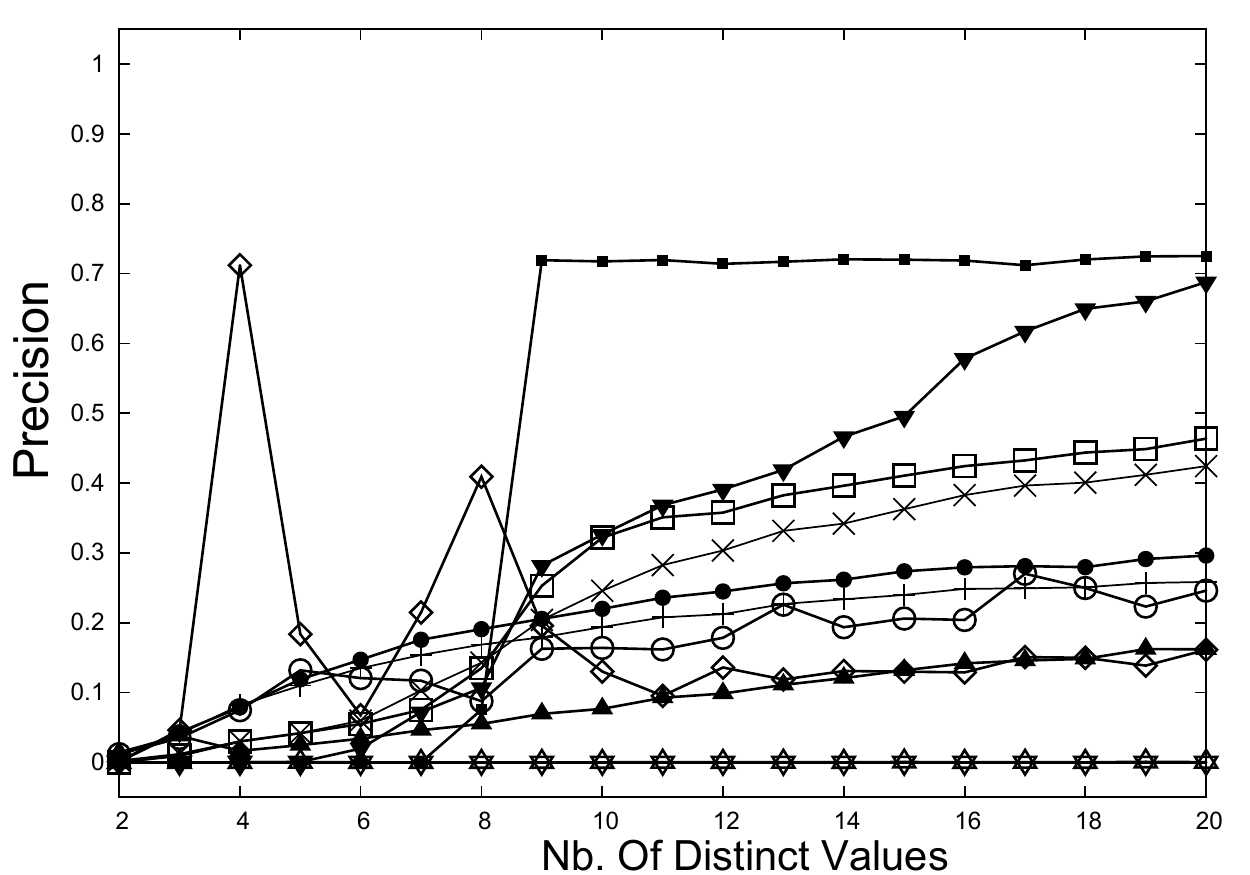}}\label{fig:E-U-FP}
\subfigure[\scriptsize{\texttt{Cov=E-Conf=U-GT=80P}}]{
\includegraphics[width=.24\linewidth]{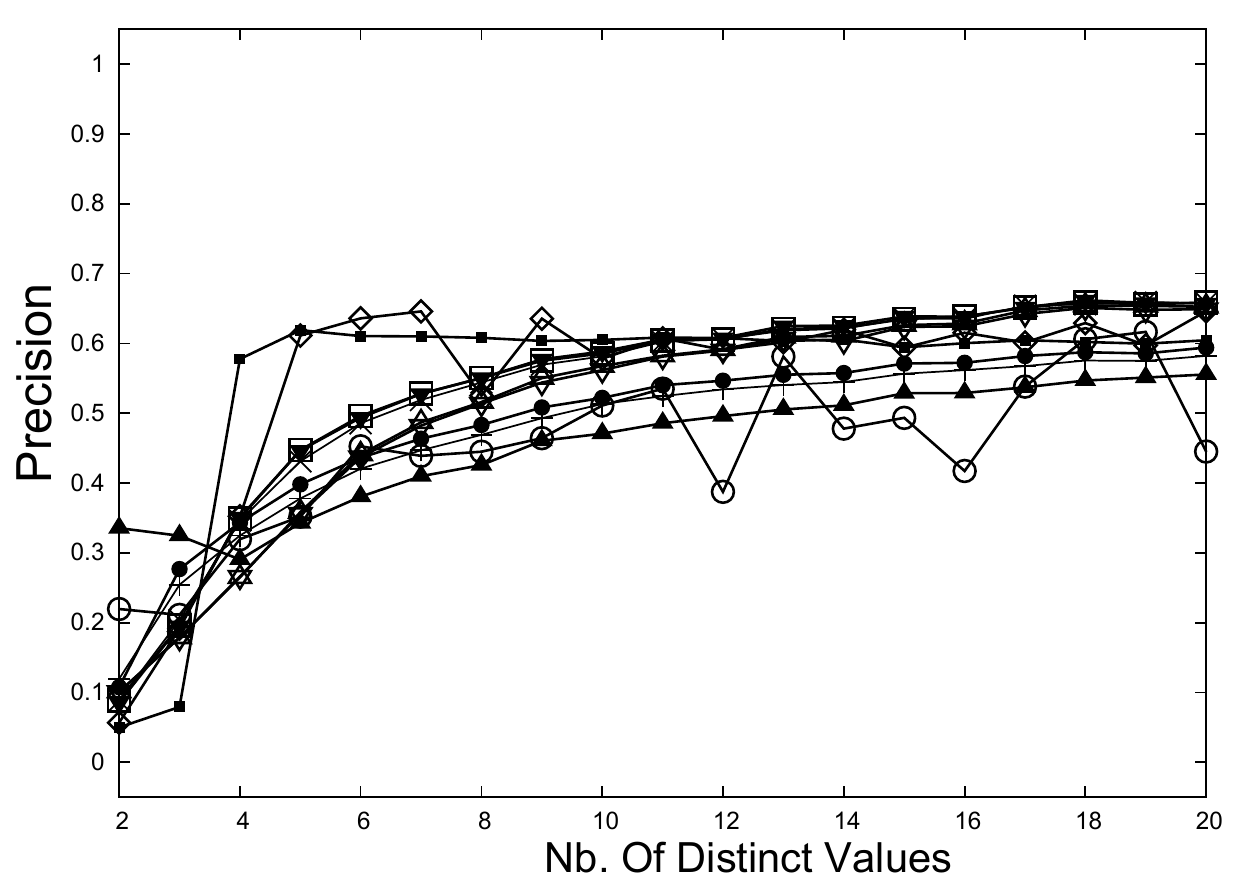}}\label{fig:EU-GT=80-P}\\
\vspace{-.2cm}
\subfigure[\scriptsize{\texttt{Cov=E-Conf=E-GT=U75}}]{
\includegraphics[width=.24\linewidth]{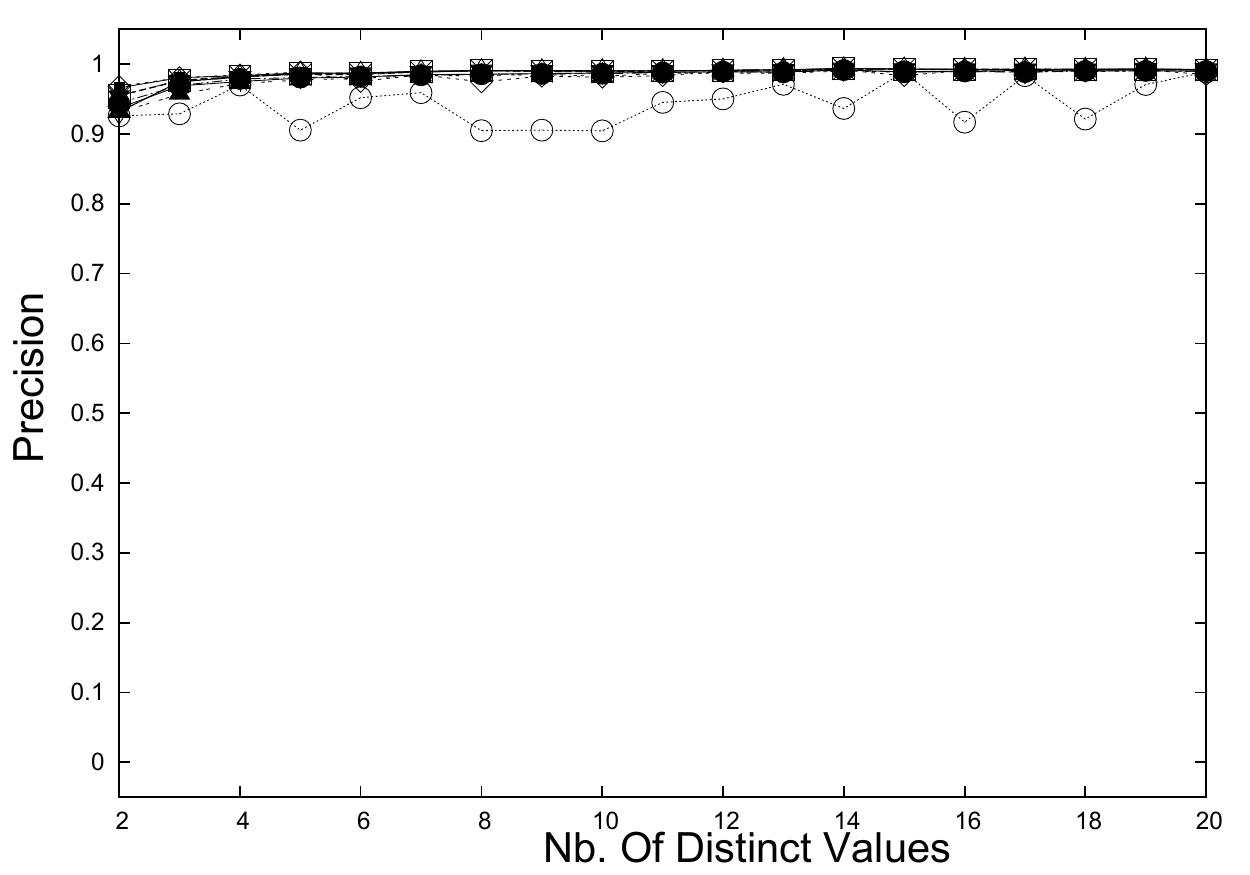}}\label{fig:E-E-GT=80-O} 
\subfigure[\scriptsize{\texttt{Cov=U25-Conf=U-GT=E}}]{
\includegraphics[width=.24\linewidth]{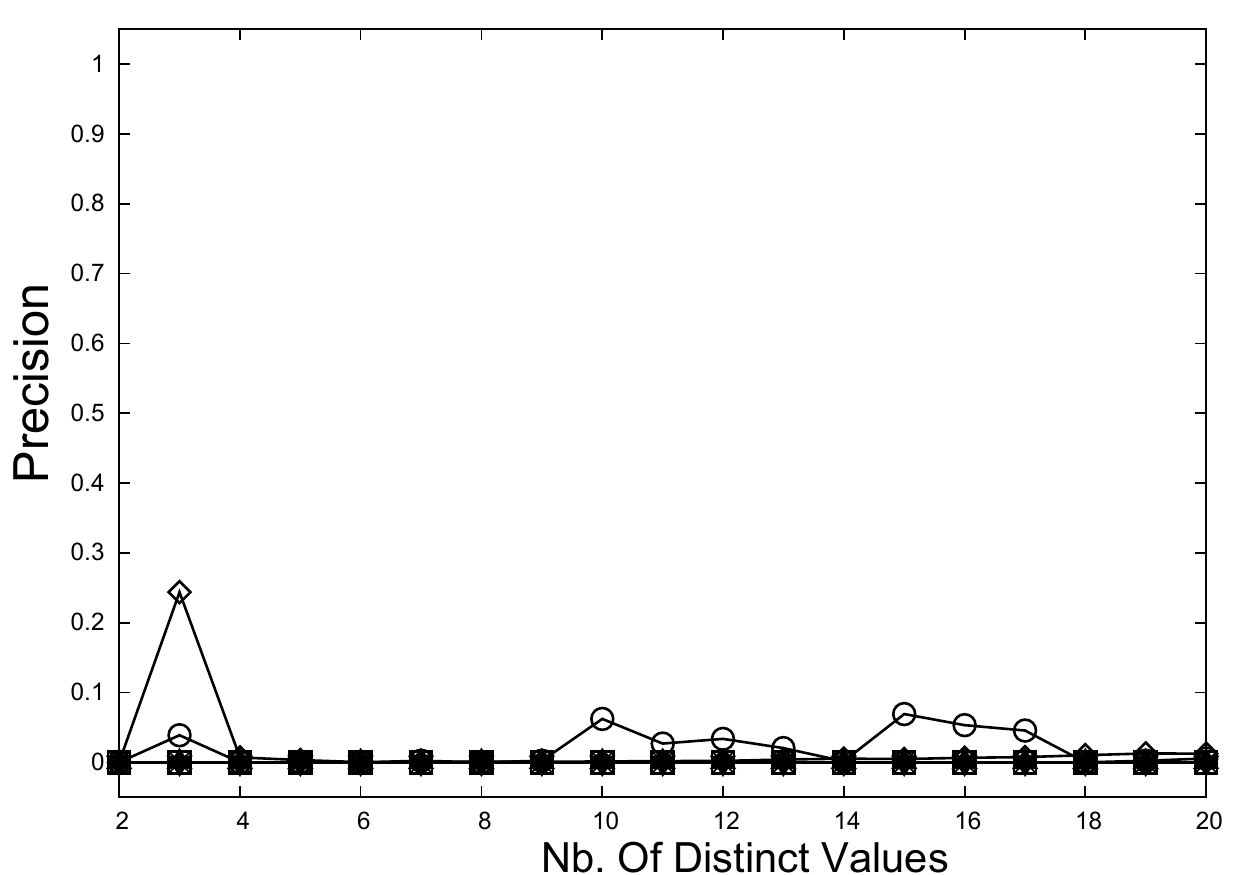}}\label{fig:U25U-GT=E} 
\subfigure[\scriptsize{\texttt{Cov=E-Conf=E-GT=FP}}]{
\includegraphics[width=.24\linewidth]{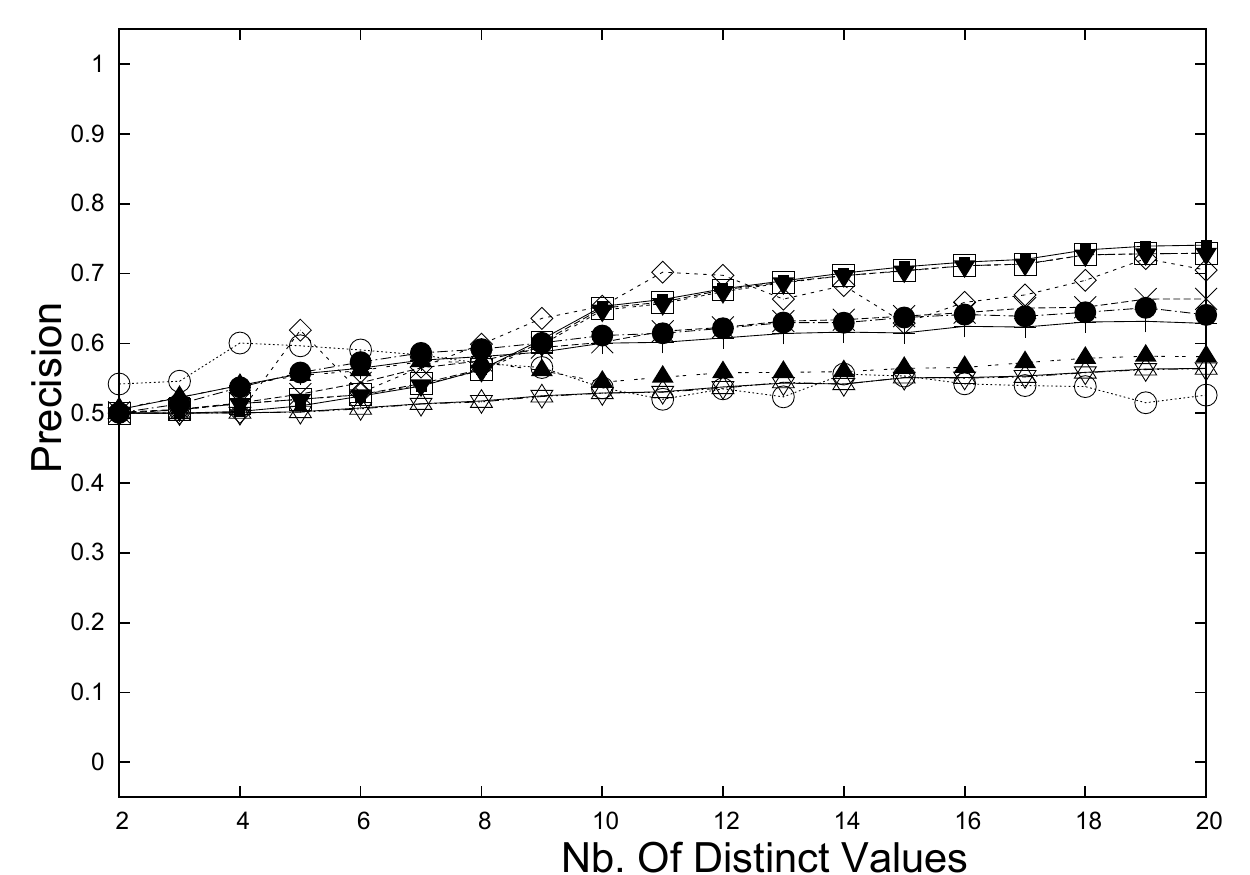}}\label{fig:E-E-GT=FP} 
\subfigure[\scriptsize{\texttt{Cov=U25-Conf=U-GT=FP}}]{
\includegraphics[width=.24\linewidth]{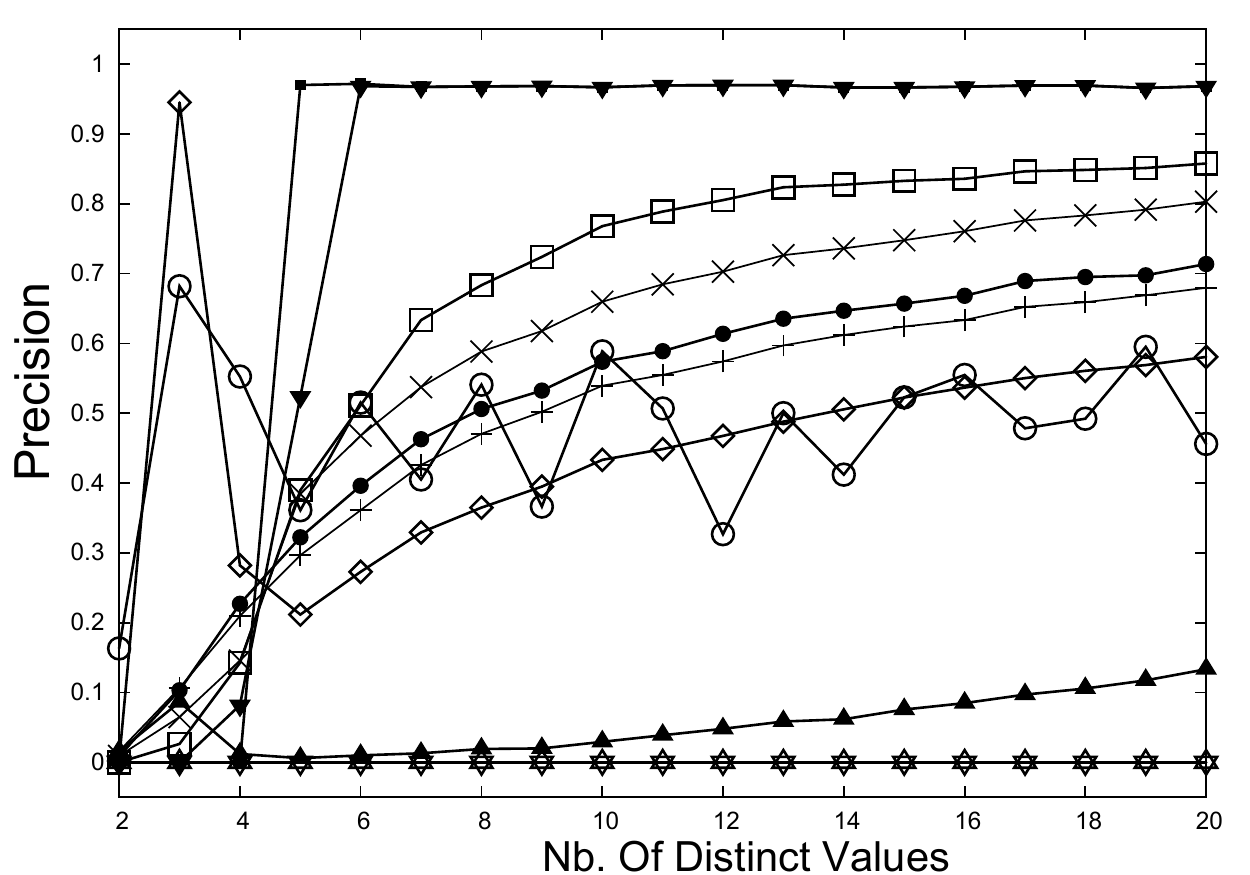}}\label{fig:U25U-GT=FP} 
\vspace{-.2cm}
\subfigure{\includegraphics[width=.4\linewidth]{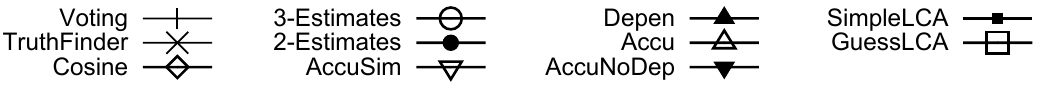}}
\caption*{{\bf Figure 1: Precision Average  for Various Truth Discovery Scenarios with $|S|=50$ and $|D|=1,000$}}
\label{fig:e}
\end{figure*}

\subsubsection{ Source Coverage}
We  compare the quality of the truth discovery models for three types of source coverage: {\scriptsize\texttt{Uniform U25}}, {\scriptsize\texttt{U75}}, and {\scriptsize\texttt{Exponential}}. Uniform source coverage corresponds to the case where all the sources provide claims for respectively 25\% or 75\% of the data items. Exponential source coverage corresponds to a more realistic case where few sources provide claims for most of the data items and the majority of the remaining sources only covers very few data items\footnote{\scriptsize{We define exponential coverage for source $i$ as: \mbox{$\forall i=0,\dots,(|S|-1), Cov_i=1+(|D|-1)\frac{e^{4i/(|S|-1)} -1  }{e^4-1}$}}}.    
 We observe that increasing the source coverage from {\scriptsize\texttt{U25}}  to {\scriptsize\texttt{U75}}  generally increases the precision of all algorithms and fewer distinct values are needed to reach the same precision, except in two cases: (1) When the distribution of true positive claims is randomly distributed across the sources  {\scriptsize\texttt{(GT=R)}}, increasing the number of data items per source does not change the precision of any method; algorithms' precision for {\scriptsize\texttt{Cov=U25}} and {\scriptsize\texttt{Cov=U75}} are identical and merged in Fig. 1(a) irrespectively of the type of conflict distribution. Precision of all methods does not differ by more than 2\% and  decreases in both cases, {\scriptsize\texttt{Conf=U}} and  {\scriptsize\texttt{Conf=E}}. (2)  When the distribution of true positive claims is exponentially distributed across the sources  {\scriptsize\texttt{(GT=E)}}, precision of all methods remains contant and close to zero even when increasing the source coverage and the number of conflicts (Fig.1 (f)).  
\subsubsection{ Conflict Distribution } 
 In the case of  exponentially distributed conflicts over the data items (Fig.~1(a) for {\scriptsize\texttt{Conf=E}} line), many data items have very few conflicts, whereas few data items have lots of conflicts\footnote{\scriptsize{We define exponential conflict distribution for data item $i$ as: $\forall i=1,\dots,|D|,
NbDistinctV_i =  (maxNbDistinctV- 1)  *\frac{e^{(2 * i / |D|)- 1}}{ e^{(2 * (|D| -1 )/|D|)-1} } + 1$}}.  Exponential conflict distribution is interesting and realistic since some data items may be more controversial and have more conflicts than others. The majority of the claims in agreement generally help all the models to reach a precision greater than .50 in the worst cases, e.g., when the sources randomly tell the truth among lots of conflicts. In that case, for {\scriptsize\texttt{Conf=E}} and {\scriptsize\texttt{GT=R}}, we observe   precision decreasing  from .75 to .525 for all methods.  For {\scriptsize\texttt{Conf=U}} and {\scriptsize\texttt{GT=R}} in Fig. 1(a),  all algorithms behave identically with decreasing precision  below .50 ({\it i.e.}, worse than random guessing). Comparing Fig. 1(c) and (g), we observe two effects when the conflict distribution type changes from uniform  to exponential: (1) precision is lifted up above .50 irrespectively of the source coverage and (2) precision range becomes more compact within .2 precision interval. 
\setcounter{subfigure}{0}
\begin{figure*}
\centering
\subfigure[$<$ 6,000 seconds]{\label{fig:Sca1}\includegraphics[height=4.2cm]{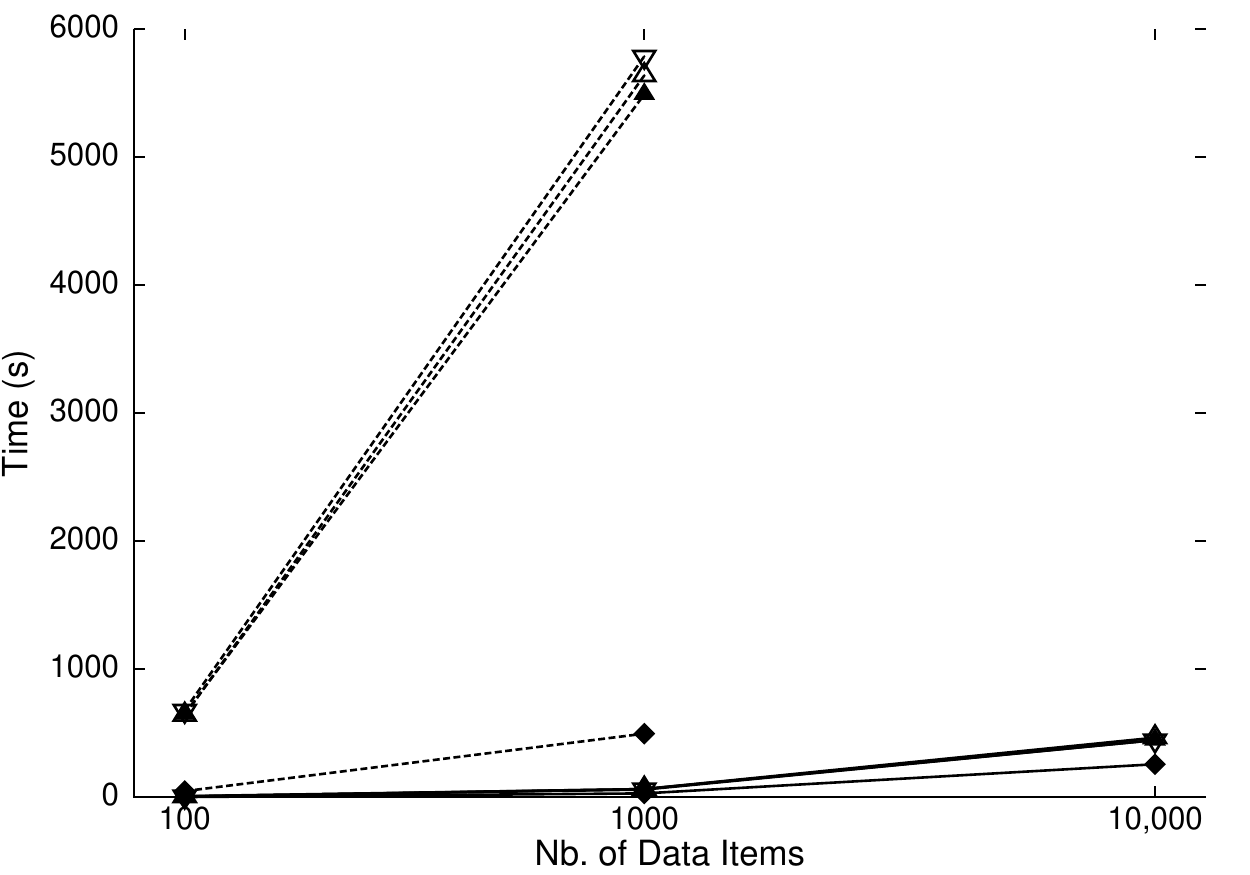}}
\subfigure[$<$ 16 seconds]{\label{fig:Sca3}\includegraphics[height=4.2cm]{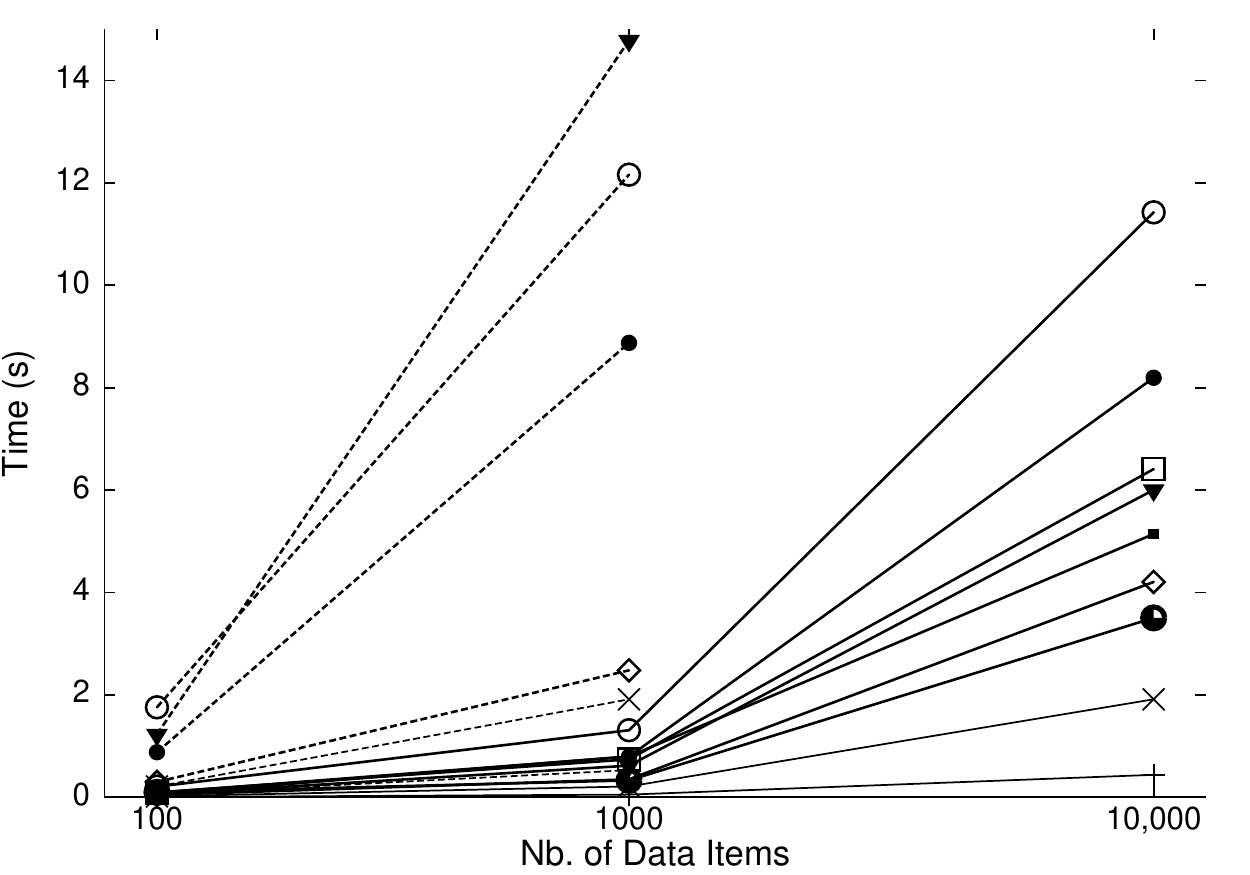}} 
\subfigure{\includegraphics[width=.1\linewidth]{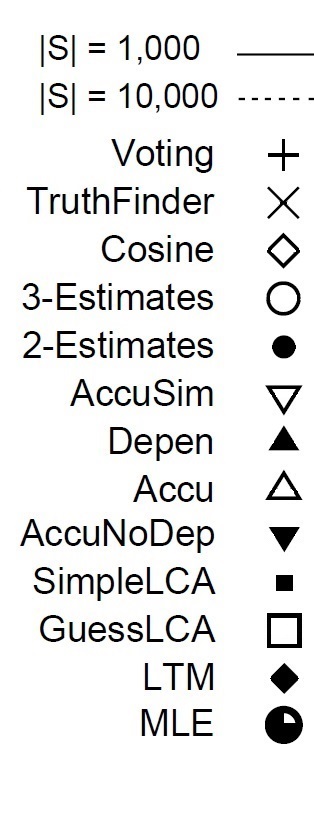}}
\caption*{{\bf Figure 2: Scalability Experiments: Runtime for scaling-up the numbers of sources  and data items}}\label{fig:Sca}
\end{figure*}

\subsubsection{Ground Truth Distribution}
Finally, we control the distribution of true positive claims among the set of claims provided by each source and we generate synthetic data sets corresponding to 7 scenarios in addition to random ({\small\texttt{GT=R}}) such as: uniform ({\small\texttt{U25, U75}}), fully pessimistic ({\small\texttt{FP}}), 80-pessimistic ({\small\texttt{80P}}), fully optimistic ({\small\texttt{FO}}), 80-optimistic ({\small\texttt{80O}}),  and exponential ({\small\texttt{E}}) as defined in Table~4.

{\bf Random  Ground Truth Distribution.}  As mentioned earlier, when true positive claims are randomly distributed across the sources, we observe that (1) none of the methods can be reliable when conflicts are uniformly distributed (decreasing precision below .50 in Fig 1(a) for {\small\texttt{Conf=U}}), and (2) increasing the source coverage or changing the  distribution of conflicts per source (from uniform to exponential) does not improve the precision of any method, (3) algorithms' precision does not differ by 2\% and decreases when the number of conflicts increases.

{\bf Uniform Ground Truth Distribution. }  
 For exponential source coverage  and  cases where the sources are equally saying the truth for 25\% of the values they claim {\small\texttt{(GT=U25)}} in Fig.~1(b), 
  the precision of the methods increases with the number of conflicts. This trend is even more significant when the source coverage increases from uniform {\small\texttt{U25}} to  {\small\texttt{U75}} since increasing the source coverage reduces the number of conflicts needed for comparable precision.  
  {\scshape 3-Estimates} has unstable results due to  the instability of $\lambda$ parameter setting. 
  In Fig. 1(b), for {\small\texttt{GT=U25}} with exponential source coverage and exponential conflict distribution, all methods behave identically and reach .75 in the best case of 20 distinct values exponentially distributed across the data items. In Fig. 1(e), when the sources are almost always, equally saying the truth {\small\texttt{(GT=U75)}}, precision of all methods does not differ more than 2\% (except  {\scshape 3-Estimates}) and reaches 1 irrespectively of the distribution type or number of conflicts. 
%
%
%
%

{\bf Pessimistic Scenarios. } In {\small\texttt{GT=FP}} scenarios of Fig. 1(c),(g), and (h), 80\% of the sources always provide false claims and 20\% always provide true claims. We observe that, for few conflicts -- {\it i.e.}, less than 8 distinct values per data item for {\small\texttt{Cov=E}} in Fig.~1(c), or 4 for {\small\texttt{Cov=U25}} in Fig.~1(h)) -- most of the methods perform worse than random guessing with precision from .1 to .4 except {\scshape Cosine} which reaches a precision peak of .7 in Fig.1(c) for 4 distinct values for {\small\texttt{Cov=E}}  and .95 precision  in Fig. 1(h) for 3 distinct values for {\small\texttt{Cov=U25}}. 
 In these two cases of source coverage, {\scshape SimpleLCA} outperforms all methods from 4 (for {\small\texttt{Cov=E}}) or 8 (for {\small\texttt{Cov=U25}}) distinct values. In Fig. 1(g), for exponential source coverage and exponential conflict distribution, precision range of the methods increases with the number of conflicts from .5 to .72. And the {\it compacting \& lifting up} effect of exponential conflict distribution on the precision of the methods is confirmed in {\small\texttt{GT=FP}} scenario of Fig. 1(g). 
   In  {\small\texttt{FP}} and  {\small\texttt{80-P}} scenarios, the  ordering of the methods based on precision remains constant: {\scshape SimpleLCA} $>$ {\scshape AccuNoDep} $>$ {\scshape GuessLCA} $>$ {\scshape TruthFinder} $>$ {\scshape 2-Estimates} $>$ {\scshape Voting} $>$ {\scshape Cosine}.  Precision of {\scshape 3-Estimates}  oscillates around or below .50  in {\small\texttt{FP}} and  {\small\texttt{80P}} scenarios in Fig.~1(c) and~(d).   {\scshape Cosine, Voting, 2-Estimates, TruthFinder}, and {\scshape GuessLCA}  behave similarly with close precision values. {\scshape Depen} and its  variants (except {\scshape AccuNoDep}) are deeply affected by random source dependence and have very low precision although increasing with the number of conflicts in Fig. 1(c) and (h).  
 For fully pessimistic scenarios with few conflicts -- either less than $4$ distinct values uniformly distributed or less than 8 distinct values exponentially distributed--  none of the methods has precision significantly better than random guessing.  In the {\small\texttt{80-Pessimistic}} scenario with exponential source coverage in Fig.~1(d), {\scshape SimpleLCA} maintains   precision greater than .55 from 4 distinct values, whereas the other methods need at least 7 distinct values to reach .50 precision. 

{\bf Optimistic Scenarios. } In {\small\texttt{GT=U75}} scenario of Fig. 1(e) with exponential source coverage and exponential conflict distribution, we observe that all methods have very similar, high precision close to 1  (except {\scshape 3-Estimates}). We  observe the same behavior with quasi-identical curves for  {\small\texttt{GT=FO}} and  {\small\texttt{GT=80O}} (see  \cite{BW2014} for detail). 
%
   In the case of optimistic scenarios with exponential source coverage and exponential conflict distribution, all methods do not differ in precision by 1\% and excel with precision close to 1 except  {\scshape 3-Estimates} which  oscillates from .9 to 1.
%
%

{\bf Exponential Ground Truth Distribution. }  This case represents the situation where one source always lies and one source always tells the truth for all the data items it covers and the remaining sources range from 1\% to 99\% of claims they provide being true\footnote{\scriptsize{We define exponential ground truth for source $i$ as: \mbox{$\forall i=1,\dots,|S|, GT_i=|D_{s_i}|\frac{e^{i/|S|} - e^{1/|S|} }{e-e^{i/|S|}}$}}}. In this case represented Fig.~1(f), none of the methods is reliable even when  the source coverage increases from  {\small\texttt{U25}}  to  {\small\texttt{U75}}. 
  None of the existing methods can cope with a wide, continuous spectrum of source truthworthiness irrespectively of the source coverage and conflict distribution, which is somehow a bad news because we can expect, in practice, that the variety of online sources  may correspond to a wide, potentially continuous range of source truthworthiness and  exponential distribution of the true positive claims per source. 

\begin{table*}[t!]
\scriptsize\centering
\begin{tabular}{|p{1.1cm}|p{4.5cm}|c|r|r|r|r|r|r|r|}
\hline {\bf Data Set}&\multicolumn{1}{c|}{{\bf Characteristics}}
&{\bf Method }& {\bf Precision }& {\bf Accuracy}& {\bf Recall}&  {\bf Specificity}& {\bf Iter.}&  \multicolumn{1}{c|}{{\bf Time}}&  \multicolumn{1}{c|}{{\bf Memory}}\\
&&& && & & &  \multicolumn{1}{c|}{(s)}& \multicolumn{1}{c|}{(MB)}\\
\hline
\multirow{10}{*}{{\bf Book} }&877 sources -- 33,235 claims&{\scshape MajorityVoting}& {\bf 0.9804}&  \textcolor{blue}{\textbf{0.8664}}& \textcolor{blue}{\textbf{0.7979}}& {\bf 0.9748}& 1&\textcolor{green}{\textbf{ 0.009}}& \textcolor{green}{\textbf{41 }}\\
& 1,263 objects&{\scshape TruthFinder}& 0.9777& 0.9387& 0.9211& 0.9667& 5& 0.359& 618\\
& 1 attribute: {\it   Author name}&{\scshape Cosine}& 0.9769& 0.9279& 0.9037& 0.9661& 8& 0.312& 165\\
& Data type: List of Strings&{\scshape2-Estimates}& 0.9812& 0.8893& 0.8351& 0.9748& 4& 0.193 & 124\\
&Gold standard count: 100 objects (7.91\%)&{\scshape 3-Estimates}& 0.9935& 0.8849& 0.8172& \textcolor{red}{\textbf{  0.9915}}& 42& 1.474& 1 117\\
&Avg coverage per source: 0.029295&{\scshape SimpleLCA}& 0.9758& 0.9023& 0.8610& 0.9667& 5& 0.136& 82\\
&Avg Nb. distinct values per data item: 3.072&{\scshape GuessLCA}& 0.9808& 0.8820& 0.8226& 0.9748& 17& 0.566& 82\\
&Avg Nb. of claims per source:37.89& {\scshape AccuSim}& 0.973& 0.9516& 0.9474& 0.9583& 3&\textcolor{blue}{\textbf{ 10.092}}& \textcolor{blue}{\textbf{2 072}} \\
&Max Nb. of distinct values per data item: 20&{\scshape Depen}& 0.9814& 0.8889& 0.8360& 0.9744& 5& 9.650& 1 879 \\
&Max Nb. of claims per source: 2 403&{\scshape Accu}& 0.9809& 0.8852& 0.8280& 0.9748& 4& 9.463& 1 451\\
&&{\scshape AccuNoDep}& 0.9806& 0.8787& 0.8172& 0.9748& 3& 0.129& 59\\
&&{\scshape LTM}& \textcolor{blue}{\textbf{0.8551}}& 0.8885&0.9839& 0.7395& 500& 4.273& 1 049\\
&& \colorbox{yellow}{{\scshape MLE}}& \laure{{\bf 1}}& \textcolor{red}{\textbf{ 1}}& \textcolor{red}{\textbf{ 1}}& \textcolor{blue}{\textbf{0}}& 2& 0.661& 590\\
\cline{3-10}
&&\textbf{ Avg}&0.9696   &	0.9060	&0.8768	  &0.8751	& 45 &2.8705&717.62\\
&&\textbf{ StdDev}&$\pm$	0.0370 &$\pm$	0.0372  &	$\pm$0.0714	&$\pm$0.2712	& $\pm137$  &$\pm$4.0711& $\pm$ 728.81 \\

\hline
\hline \multirow{8}{*}{{\bf Flight} }& 38 sources -- 2,864,985 claims&{\scshape MajorityVoting}&{\bf 0.8205} & {\bf 0.8228} & {\bf 0.8199} &{\bf  0.8256} & 1 & \textcolor{green}{\textbf{0.485}} & \textcolor{green}{\textbf{274}} \\
&34,652 objects -- 207,908 data items &{\scshape TruthFinder}& 0.7970 & 0.7997 & 0.7965 & 0.8028 & 2 & 3.974 & 673 \\
&6 attributes: {\it Expected/Actual Departure}&{\scshape Cosine} & 0.8825 & 0.8839 & 0.8819 & 0.8859 & 13 & 42.696 & 1 366\\
&\hspace{.4cm}{\it /Arrival Time/Gate}&{\scshape 2-Estimates}& 0.7903 & 0.7931 & 0.7898 & 0.7963 & 4 & 17.444 & 1 413\\
&Data type: (String,Time)&{\scshape 3-Estimates}& 0.7028 & 0.7068 & 0.7023 & 0.7112 & 24 & 92.020 & 1 622 \\
&Gold standard count: 16,134 values (7.76\%)&{\scshape SimpleLCA}& \textcolor{blue}{\textbf{0.6802}}& \textcolor{blue}{\textbf{0.6846}} & \textcolor{blue}{\textbf{0.6797}}& \textcolor{blue}{\textbf{0.6893 }}& 7 & 7.904 & 1 612\\
&Avg coverage per source: 0.36263&{\scshape GuessLCA}& 0.7867 & 0.7895 & 0.7861 & 0.7927 & 137 &\textcolor{blue}{\textbf{417.289}}& 1 606 \\
&Avg Nb. distinct values per data item: 2.2783&{\scshape AccuSim}& 0.9049 & 0.9059 & 0.9042 & 0.9076 & 4 & 65.288 & 1 623 \\
&Avg Nb. of claims per source: 75,394.34&{\scshape Depen}& 0.8204 & 0.8227 & 0.8198 & 0.8255 & 2 & 53.261 & 1 622 \\
&Max Nb. of distinct values per data item: 14& \colorbox{yellow}{{\scshape Accu}}& \laure{\textbf{0.9111}} & \textcolor{red}{\textbf{0.9121}} & \textcolor{red}{\textbf{0.9105}} & \textcolor{red}{\textbf{0.9136}} & 3 & 58.550 & \textcolor{blue}{\textbf{1 629}} \\
&Max Nb. of claims per source: 197 103&{\scshape AccuNoDep}& 0.7915 & 0.7942 & 0.791 & 0.7974 & 3 & 13.173 & 1 237\\
\cline{3-10}
&&\textbf{ Avg}& 0.8080&0.8105&0.8074&0.8134& 18&70.1895&1 334.27\\
&&\textbf{ StdDev}&	$\pm$0.0738&	$\pm$0.0727&	$\pm$0.0737&	$\pm$0.0717&$\pm 40$&$\pm$	118.8748& $\pm$454.51\\

\hline
\hline \multirow{8}{*}{{\bf Weather}}& 16 sources -- 365,890 claims 
&{\scshape MajorityVoting}&{\bf 0.6305} & {\bf 0.8472 }& {\bf 0.649 }& {\bf 0.8995} & 1 & \textcolor{green}{\textbf{0.089}} & \textcolor{green}{\textbf{24}} \\
 &6,375 objects -- 30,317 data items & \colorbox{yellow}{{\scshape TruthFinder}}& \laure{\textbf{0.6443}} & \textcolor{red}{\textbf{0.8531}} & \textcolor{red}{\textbf{0.6633}} & \textcolor{red}{\textbf{0.9033}} & 2 & 1.238 & 476 \\
&5 attributes: {\it Temperature, Real Feel, } &{\scshape Cosine} & 0.6283 & 0.8462 & 0.6468 & 0.8989 & 9 &5.405 & 2 348 \\
&\hspace{.4cm}{\it  Humidity, Pressure, Visibility}&{\scshape 2-Estimates}& 0.6310 & 0.8474 & 0.6495 & 0.8996 & 5 & 3.417 & 1 823\\
&Data type: Number&{\scshape 3-Estimates}&  0.6272 & 0.8457 & 0.6456 & 0.8986 & 6 & 5.261 &\textcolor{blue}{\textbf{2 354}} \\
&Gold standard count: 22,570 values (74.4\%)&{\scshape SimpleLCA}& 0.6421 & 0.8522 & 0.6610 & 0.9027 & 4 & 1.687 & 1 009\\
&Avg coverage per source: 0.754&{\scshape GuessLCA}& 0.6359 & 0.8495 & 0.6546 & 0.9010 & 11 &\textcolor{blue}{\textbf{6.741} }& 1 346  \\
&Avg Nb. distinct values per data item: 4.546&{\scshape AccuSim}& \textcolor{blue}{\textbf{0.5079}}&\textcolor{blue}{\textbf{0.7944}} & \textcolor{blue}{\textbf{0.5229}} &\textcolor{blue}{\textbf{0.8662} }& 3 & 4.610 & 2 259 \\
&Avg Nb. of claims per source: 22,868.12&{\scshape Depen}& 0.6305 & 0.8472 & 0.6490 & 0.8995 & 3 & 4.284 & 2 157 \\
&Max Nb. distinct values per data item: 17&{\scshape Accu}& 0.5231 & 0.8010 & 0.5385 & 0.8703 & 3 & 4.332 & 2 206\\
&Max Nb. of claims per source: 29 290&{\scshape AccuNoDep} & 0.6442 & \textcolor{red}{\textbf{0.8531}} & 0.6631 & 0.9032 & 3 & 2.451 & 1 127\\
\cline{3-10}
&&\textbf{ Avg}&0.6132&0.8397&0.6312&0.8948& 5&3.5923&1557.18\\
&&\textbf{ StdDev}&$\pm$ 0.0488	&$\pm$0.0210	&$\pm$0.0502&	$\pm$0.0133&	$\pm 3$&$\pm$2.0205&$\pm$815.72 \\
\hline
\hline
\multirow{8}{*}{{\bf Population}}&4,264 sources -- 49,955 claims
& {\scshape MajorityVoting}& {\bf 0.8206 }& {\bf 0.8419} &{\bf  0.8373} & {\bf 0.8457} & 1 & \textcolor{green}{\textbf{0.044}} & \textcolor{green}{\textbf{19}} \\
& 41,196 objects -- 42,832 data items&{\scshape TruthFinder}& 0.8505 & 0.8698 & 0.8678 & 0.8714 & 2 & 0.349 & 60\\
& 1 attribute: {\it City Population per year}&{\scshape Cosine}& 0.8306 & 0.8512 & 0.8475 & 0.8543 & 7 & 0.629 & 120\\
& Data type: Number&{\scshape 2-Estimates}& \textcolor{blue}{\textbf{0.6777}} & \textcolor{blue}{\textbf{0.7085}} & \textcolor{blue}{\textbf{0.6915}} & \textcolor{blue}{\textbf{0.7229}} & 6 & 0.835 & 260\\
&Gold standard count: 301 values (0.702\%)&{\scshape 3-Estimates}& 0.8239 & 0.8450 & 0.8407 & 0.8486 & 8 & 1.178 & 300\\
&Avg coverage per source: 2.67E-4&{\scshape SimpleLCA}& 0.8372 & 0.8574 & 0.8542 & 0.8600 & 4 & 0.343 & 120\\
&Avg Nb. distinct values per data item: 1.041&{\scshape GuessLCA}& 0.8239 & 0.8450 & 0.8407 & 0.8486 & 5 & 0.691 & 160\\
&Avg Nb. of claims per source: 11.715& {\scshape AccuSim}& 0.8206 & 0.8419 & 0.8373 & 0.8457 & 5 & 101.174 &\textcolor{blue}{\textbf{1 625}} \\
&Max Nb. distinct values per data item: 11&{\scshape Depen}& 0.8173 & 0.8388 & 0.8339 & 0.8429 & 4 & 101.078 & 699 \\
&Max Nb. of claims per source: 25 820&{\scshape Accu}& 0.8472 & 0.8667 & 0.8644 & 0.8686 & 4 & \textcolor{blue}{\textbf{106.336}} & 499\\
&& \colorbox{yellow}{{\scshape AccuNoDep}}& \laure{\textbf{0.8538}} & \textcolor{red}{\textbf{0.8729}} & \textcolor{red}{\textbf{0.8712}} & \textcolor{red}{\textbf{0.8743}} & 4 & 0.411& 174\\
\cline{3-10}
&&\textbf{ Avg}& 0.8185&0.8399&0.8351&0.8439& 5&28.46076&366.91\\
&&\textbf{ StdDev}&$\pm$ 0.0485&$\pm$	0.0452	&$\pm$0.0494	&$\pm$0.0417	&$\pm 2$&$\pm$47.8053&$\pm$462.99\\
\hline\hline
\multirow{8}{*}{{\bf Biography}}&771,132 sources -- 10,862,648 claims
& {\scshape MajorityVoting}& {\bf 0.7068} & {\bf 0.8961 }& {\bf 0.9032 }& {\bf 0.8941} & 1 & \textcolor{green}{\textbf{3.342}} & \textcolor{green}{\textbf{1439}} \\
& 1,863,248 objects -- 3,783,555 data items&{\scshape TruthFinder}& 0.7064 & 0.8959 & 0.9027 & 0.8940 & 4 & 44.515 & 7 487\\
& 9 attributes: {\it Born, Died, Spouse, Father,  } &{\scshape Cosine}& \textcolor{blue}{\textbf{0.7037}}&\textcolor{blue}{\textbf{0.8944}}&\textcolor{blue}{\textbf{0.8993}}& \textcolor{blue}{\textbf{0.8930}} & 2 & 32.599 & 7 470\\
& {\it Mother, Children, Country, Height, Weight}& \colorbox{yellow}{{\scshape 2-Estimates}}& \laure{ \textbf{0.7091}}& \textcolor{red}{\textbf{0.9409}} & \textcolor{red}{\textbf{0.9061}} & \textcolor{red}{\textbf{0.8950}} & 2 & 32.979 & 7 470\\
&Data type: (List of names, Date, Numerical)&{\scshape 3-Estimates}& 0.7060 & 0.8957 & 0.9022 & 0.8939 & 24 & \textcolor{blue}{\textbf{317.305}}& \textcolor{blue}{\textbf{8 771}}\\
&Gold standard count: 2,626 values (0.069\%)&{\scshape SimpleLCA}& NA & NA & NA & NA & 500 &NA &  NA\\
&Avg coverage per source: 3.72E-6&{\scshape GuessLCA}& NA&NA& NA & NA & 500 & NA & NA\\
&Avg Nb. distinct values per data item: 1.05&{\scshape  AccuSim}& EL & EL & EL & EL & EL & EL & EL \\
&Avg Nb. of claims per source: 14.08&{\scshape Depen}& EL & EL & EL & EL & EL & EL & EL  \\
&Max Nb. of conflicts: 60&{\scshape Accu}& EL & EL & EL & EL & EL & EL & EL\\
&Max Nb. of claims per source: 2 839 091&{\scshape AccuNoDep}& 0.7053 & 0.8953 & 0.9012 & 0.8936 & 3 & 80.924 & 7 488\\
\cline{3-10}
&&\textbf{ Avg}&0.7062&0.8958&0.9025&0.8939&5 &85.2774&	6 687.66\\
&&\textbf{ StdDev}& 	$\pm$0.0018&	$\pm$0.001	&$\pm$0.0023	&$\pm$0.0007&$\pm 8$&	$\pm$116.39&$\pm$2 622.74\\
\hline
\end{tabular}
\caption*{{\bf Table 5. Experimental Results for Real-World Data Sets}}\label{tab:real-world dataset}
\end{table*}

\setcounter{subfigure}{0}
\begin{figure*}
\centering
\subfigure[{\scriptsize Book}]{\label{fig:book}
\includegraphics[width=3.3cm,height=2.5cm]{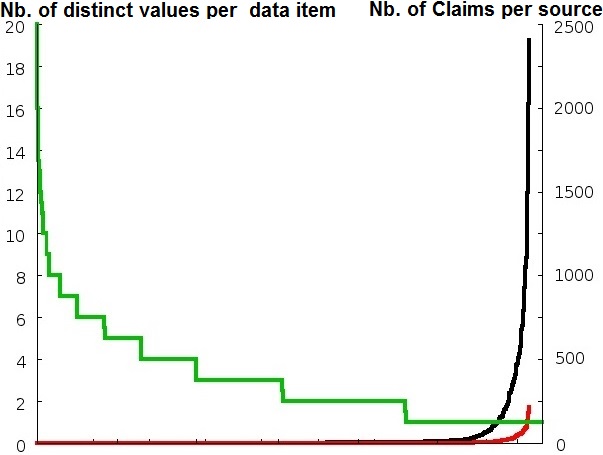}}
\subfigure[{\scriptsize Flight}]{\label{fig:flight}
\includegraphics[width=3.3cm,height=2.5cm]{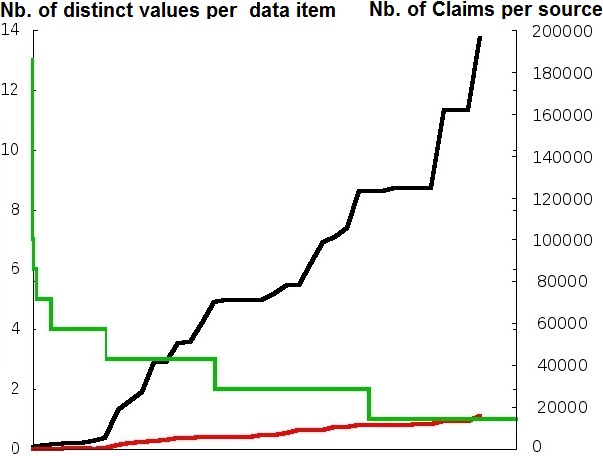}}
\subfigure[{\scriptsize Weather} ]{\label{fig:weather}
\includegraphics[width=3.3cm,height=2.5cm]{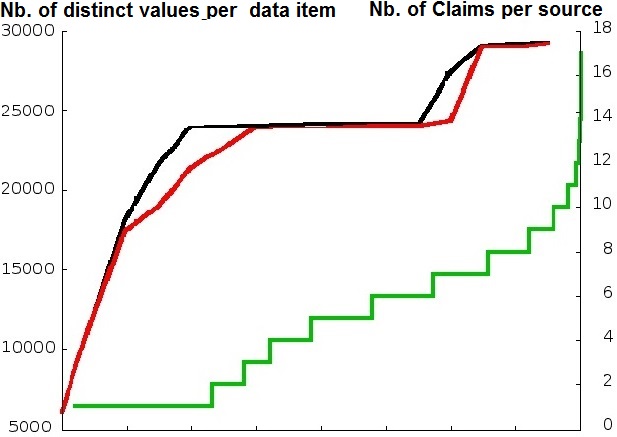}}
\subfigure[{\scriptsize Population}]{\label{fig:population} 
\includegraphics[width=3.3cm,height=2.5cm]{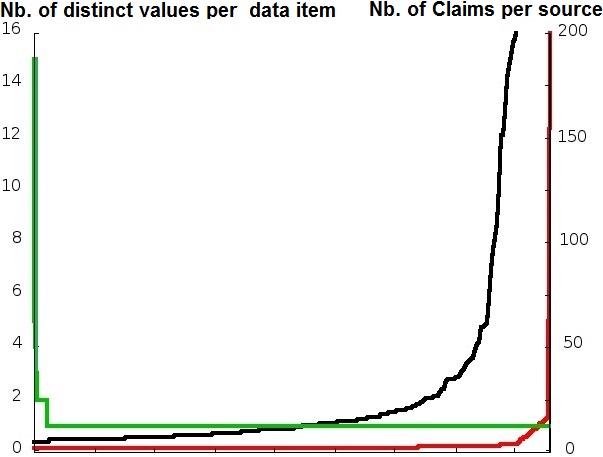}}
\subfigure[{\scriptsize Biography }]{\label{fig:biography}
\includegraphics[width=3.3cm,height=2.5cm]{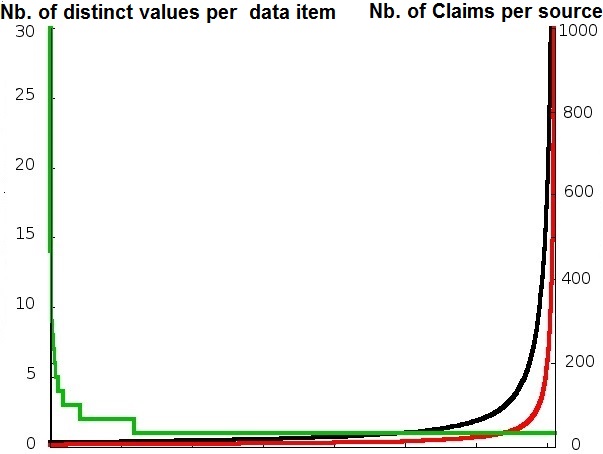}} \\
\vspace{-.2cm}
\subfigure{\includegraphics[width=.63\linewidth,height=.28cm]{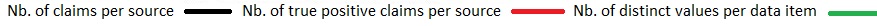}}
\caption*{{\bf Figure 3: Distribution of Claims and Positive True Claims per Source and Conflicts per Data Items in the Real-World Data Sets}}
\label{fig:rw-distributions}
\end{figure*}
%

\subsection{Scalability Experiments}
To characterize the different algorithms' behavior in terms of scalability, we evaluate them using large synthetic data sets. Each reported time is the average of 10 executions over 10 different data sets of the same size and same configuration as {\small\texttt{Cov=U25-Conf=U-GT=R}} for which all methods obtain the same precision. We increased the number of data items from 100 to 10,000 and the number of sources from 1,000 to  10,000. The experiment with 10,000 sources and 10,000 data items ({\it i.e.}, 100 millions claims) exceeded our main memory capacity and is not reported. 

Let {\it Sca}$_D$ be the case with $|S|=1,000$ sources and $|D|=10,000$  data items. Let {\it Sca}$_S$ be the case with $|S|=10,000$ sources and $|D|=1,000$  data items.  Fig.~2 shows two types of runtime behavior. Fig.~2(a) presents the models including source dependence computation ($<$ 6,000 seconds). Fig.~2(b) presents the other algorithms ($<$ 16 seconds). LTM lies between these two types of behavior with 256 seconds for {\it Sca}$_D$ and twice more (496 seconds) for {\it Sca}$_S$ and it is plotted in Fig~2(a).  
   For a large number of sources ($|S| > 5,000$), the time for MLE and LCA models could not be reported since the algorithms obtained $0/0$ undetermined form (NaN) for the value confidence and source truthworthiness computation.  In Fig.~2(a),  {\scshape Depen}, {\scshape Accu}, and {\scshape AccuSim} exhibit similar performance of linear scaling on the number of data items for 1,000 sources ({\it Sca}$_D$ in solid lines),  
but quadratic scaling on the number of sources  ({\it Sca}$_S$ in dashed lines): from 5,492 seconds for {\scshape Depen} to 5,788 seconds for {\scshape AccuSim}. 
   Fig.~2(b) shows the fastest algorithms with runtime below 12 seconds for {\it Sca}$_D$ and below 16 seconds for {\it Sca}$_S$.     For {\it Sca}$_D$,  MLE performs faster than {\scshape Cosine} and LCA models. 2- and {\scshape 3-Estimates} are the slowest but they maintain almost the same execution times in the two cases, slightly lower for {\it Sca}$_S$.  For {\it Sca}$_S$, {\scshape AccuNoDep}  is the slowest algorithm after  2- and {\scshape 3-Estimates}.  
 These results corroborate the time complexity analysis given in Table~3. Finally, Fig.~2(b) demonstrates the efficiency of {\scshape MajorityVoting} and {\scshape TruthFinder} in both cases:  
  438 and 528 milliseconds for {\scshape MajorityVoting} for {\it Sca}$_D$ and {\it Sca}$_S$ respectively and   1.912 seconds for  {\scshape TruthFinder} in both cases. 
  
   From our scalability experiments, we can conclude that {\scshape MajorityVoting} and {\scshape TruthFinder} perform best for truth discovery on our synthetic  data sets. This concerns both the scaling on the number of data items and claims, as well as the scaling on the number of sources. 

\setcounter{subfigure}{0}
\begin{figure*}
\centering
\subfigure[{\small Book Data and Optimistic Scenarios}]{\includegraphics[height=4.1cm]{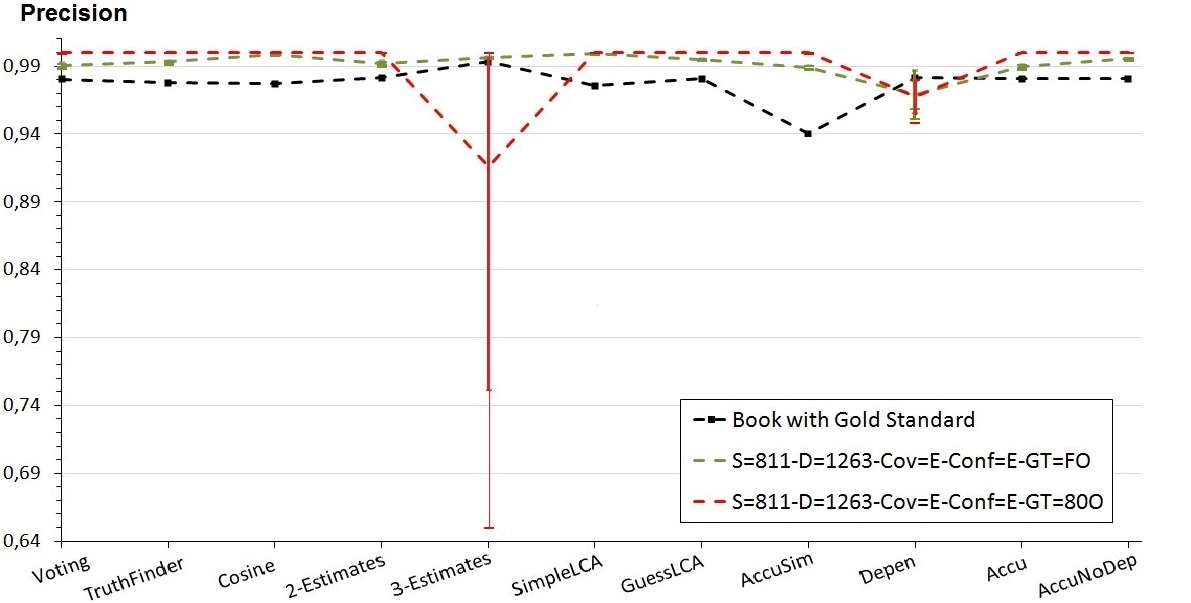}}
\subfigure[{\small Weather Data and Pessimistic Scenario}]{\includegraphics[height=3.9cm]{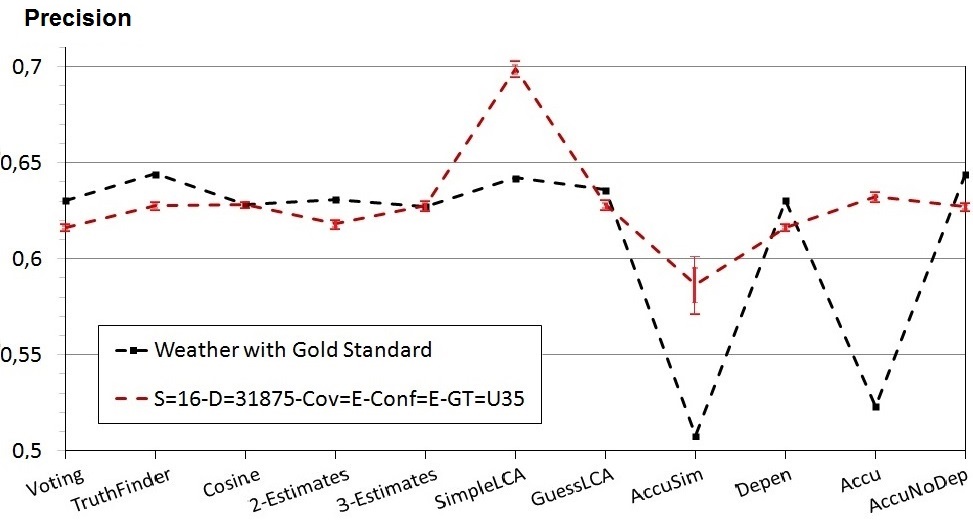}}

\caption*{{\bf Figure 4: Comparison of Algorithms' Precision for Real-World and Synthetic Data sets} }\label{fig:RS}
\end{figure*}

%

\subsection{Experiments on Real-World Data}

In this set of experiments, our goal is to compare quality metrics and performance of the algorithms on five real-world data sets. 
Table~5 shows the characteristics of these data sets and provides the quality metrics, number of iterations,  execution time, and memory usage (with EL when exceeding the time or memory capacity limits of the experiments and NA when value confidence calculation underflows to zero and  source truthworthiness computation produces NaN result). In Table~5, red color indicates the best quality metrics, yellow highlight the winner based on maximal precision, green indicates the fastest execution  and lowest memory consumption, whereas blue indicates the worst quality metrics, the slowest execution and the highest memory consumption. Results of {\scshape MajorityVoting} as the baseline are in black bold. Fig. 3 shows the distributions of claims per source (black line),  true positive claims  per source (red line), and  distinct values per data items (green line).

\noindent{\bf Book.} The Book data set from \cite{Dong:2010} 
 consists of 33,235 claims on the author names of 1,263 books by 877 book seller sources. The gold standard consists of 100 
randomly sampled books for which the book covers were manually verified by the authors of \cite{Dong:2010} representing 100/1263 = 7.91\% of the complete ground truth. Distributions are illustrated in Fig. 3(a). 
A version of the Book data set has been formatted so that MLE could be compared. MLE reaches precision 1, accuracy 1, recall 1, and null specificity in 2 iterations and 661 milliseconds. It outperforms all methods including {\scshape MajorityVoting} when we compare the  gain in precision versus the loss in execution time. {\scshape 3-Estimates} is ranked in the second position for precision but first for specificity: this can be explained by the optimal tuning of its parameters for the Book data set. {\scshape Depen} models have the third position in terms of precision but expose prohibitive runtime due to source dependence computation. Even after 500 iterations, LTM has the lowest precision.

\noindent{\bf Flight.} The Flight data set from \cite{LiDLMS12} consists of 2,864,985 claims from 38 sources on 34,652 flights for 6 attributes with distributions illustrated in Fig. 3(b). 
  The gold standard contained 16,134 true values which represents 7.76\% of the complete ground truth. {\scshape Accu} outperforms all methods for all quality metrics with the highest memory consumption but a reasonable runtime for 3 iterations compared to the average and the worse case of {\scshape GuessLCA} in terms of time and number of iterations. However, {\scshape Accu} is about 120 times slower than {\scshape MajorityVoting} for only +.0906 precision increase.

\noindent{\bf Weather.}  The Weather data set from \cite{Dong:2010}  consists of 426,360 claims from 18 sources on the Web for 5 attributes 
  on hourly weather predictions for 49 US cities between January and February 2010 (Fig. 3(c)). As gold standard, we used 
 30,170 claims from AccuWeather Web site which can cover 74.4\% of the complete ground truth.  {\scshape AccuSim} and {\scshape Accu} are penalized mainly because weather data are very similar by nature and the weight on similarity is misleading: they did not perform better than random guessing when sources make lots of false claims. However, {\scshape TruthFinder} is the winner reaching .6443 precision after  1,238 milliseconds and 2 iterations, only 13 times slower than {\scshape MajorityVoting} with +.0138 precision increase (Table 4). Again {\scshape GuessLCA} is the slowest almost doubling the average time in 11 iterations.

\noindent{\bf Population.}  The Population data set from \cite{Pasternack:2010} consists of 49,955 claims extracted from Wikipedia edits from 4,264 sources  (Fig. 3(d)). The gold standard used by the authors was 301 true values on the population from US Census representing .702\% of the complete ground truth.  {\scshape AccuNoDep} outperforms all methods in 411 milliseconds and 4 iterations, 9 times slower than  {\scshape MajorityVoting} for +.0332 precision increase. {\scshape Accu} is the slowest and {\scshape AccuSim} has maximal memory consumption due to similarity computation.

\noindent{\bf Biography.} We extended the Biography data set extracted from Wikipedia in \cite{Pasternack:2010}  with 10,862,648 claims over 19,606 people and 9 attributes 
  from 771,132 online sources (Fig. 3(e)). The gold standard consists in 2,626 true values from authoritative sources representing .069\% of the complete ground truth. Computing source dependence expose a prohibitive runtime (EL) and confidence computation by LCA models was not feasible (NA). Finally, 2-{\scshape Estimates} has the best quality metrics in only 2 iterations but almost 10 times slower than {\scshape MajorityVoting} for  5 times more memory usage and +.0023 precision increase.

From Fig. 2 and Table 5, we observe that all real-world data sets have exponential source coverage {\small\texttt{(Cov=E)}} 
   and  exponential distribution of their distinct values {\small\texttt{(Conf=E)}}. To confront our findings from the experiments on synthetic data, we generate data sets mimicking the characteristics of the Book and Weather data sets in Fig. 4 with the advantage to generate the {\it complete ground truth}.
   
  {\bf Optimistic scenarios.}     The sources of the Book data set generally have no interest in providing wrong information about their products and we can assume that their underlying ground truth distribution can either be {\small\texttt{80-Optimistic}} or {\small\texttt{Fully Optimistic}} in the best case. Fig.~4(a) presents the precision of the algorithms for the Book data set with its original gold standard, as well as the averaged precision over 10 synthetic data sets generated with similar characteristics in terms of numbers of sources and data items  for {\small\texttt{80O}} and {\small\texttt{FO}} scenarios with maximum 20 conflicts exponentially distributed. All methods have very high precision for the optimistic scenarios with precision in the following 95\% confidence intervals: [.9897;.9934] for {\small\texttt{GT=FO}} and [.9732;1] for {\small\texttt{GT=80O}}  over the total number of true positive claims we generated. In that case, we can conclude that  the results obtained from the synthetic data with complete ground truth corroborate the ones obtained from the gold standard of the Book data set. This gold standard has been carefully selected and we observe that it can be considered as a representative sample of the complete ground truth.
   
    {\bf Pessimistic scenarios.} In Table 5, precision average for the Weather data set is ($.6134 \pm .0489$), computed from a gold standard that was considered as an authoritative source. We generated many data sets with similar characteristics in terms of number of sources and data items, numbers and distributions of claims per source and distinct values per data items for a wide range of pessimistic scenarios. Fig. 4(b) represents the closest precision we could get for {\small\texttt{GT=U35}}. We can observe that precision obtained for the gold standard with 74.4\% of the original data set size actually corresponds to the precision we can obtain with synthetic data sets generated for a scenario where 35\% of the total number of claims provided by the sources uniformly are true positive claims. This leads us to put into perspective the authoritativeness of AccuWeather source as a gold standard, despite its coverage.

 Finally, we observe that none of the considered algorithms has clear benefits over {\scshape MajorityVoting} when we compare the gain in precision of the best method ($+.0319 \pm 0.0696$) versus the loss in runtime ($+17.97 \pm 58.67$) seconds in average for the five real-world data sets.  Moreover,  experiments on real-world data sets confirm our observations: the algorithms of our study have been originally designed to excel in {\it optimistic} scenarios with lots of conflicts (from maximum 11 to 60) exponentially distributed across all data items. For data sets where most of the sources provide false claims still with lots of conflicts, the methods precision is relatively low (from .6134 to .7072 in average).  
  The experimental results obtained from real-world data sets corroborate the results we obtained from the experiments on the synthetic data sets and demonstrate that our framework and data set generator can help in cross-checking data set gold standards.

\section{Conclusions}~\label{sect:conclusions}
Reimplementing and extensively comparing 12 algorithms for truth discovery from multi-source, conflicting data was a challenging task, mainly due to the problems we faced to set up the experimental framework to compare all methods in a unified and fair way. Even so, we had to omit other existing algorithms  
related to source trust assessment \cite{VydiswaranZR11}, Web link analysis 
 \cite{Pasternack:2010}, and recent work on correlated data~\cite{Pochampally2014} and conflict resolution~\cite{sigmod14}.   
 Our main conclusions are the following:  (1) Stability and repeatability of the results are significant issues for {\scshape LTM} and {\scshape 3-Estimates}. Fluctuations of their results are due to randomization in {\scshape LTM} and normalization in {\scshape 3-Estimates}. Multiple executions of these algorithms are required to compute meaningful averages of the quality metrics. We also observed that parameter setting can dramatically impact the quality of these algorithms.  (2) When the number of sources exceeds 5,000: LCA and MLE computation is not feasible (0/0) or exceeds the memory capacity limit for {\scshape Depen}, {\scshape Accu}, and {\scshape AccuSim} models. 
  (3) All methods do not perform significantly better than random guessing when the data set has few conflicts per data item and a large number of non reliable sources (pessimistic scenarios). (4) Although {\scshape MajorityVoting} can be misleading when sources are dependent, it remains the most efficient and scalable for a minor degradation in precision compared to  the other methods that are from 9 ({\scshape TruthFinder}) to 120 times ({\scshape Accu}) slower.

   Future work consists of extending this work in a number of fronts. Firstly, we hope that our synthetic data set generation framework can be used and extended for parameter setting, testing  and in-depth evaluation of other existing or new algorithms in a variety of truth discovery scenarios (e.g., with controlling source dependence and value  similarity). The main advantage of our framework is to control a complete ground truth (usually hard to get with real-world data sets) and mimic real-world truth discovery scenarios.  Secondly, we can see many challenging research avenues for the next generation of truth discovery methods: (1) To improve scalability on the number of sources to be applicable to data from social networks and social media, (2) To improve the algorithms' precision for pessimistic scenarios when most of sources are not reliable and have few conflicting values, (3) To improve the usability and repeatability of the algorithms, either by simplifying the parameterization or combining multiple methods to find optimal parameter setting.

\balance

\begin{scriptsize}
\bibliographystyle{abbrv}
\bibliography{references} 

\begin{thebibliography}{10}

\bibitem{BalakrishnanK11}
R.~Balakrishnan and S.~Kambhampati.
\newblock {SourceRank: Relevance and Trust Assessment for Deep Web Sources
  based on Inter-source Agreement}.
\newblock In {\em WWW}, pages 227--236, 2011.

\bibitem{BW2014}
L.~Berti-Equille and D.~A. Waguih.
\newblock {Truth Discovery Algorithms: An Experimental Evaluation, QCRI
  Technical Report, May 2014}, 2014.

\bibitem{DongBHS10a}
X.~Dong, L.~Berti-Equille, Y.~Hu, and D.~Srivastava.
\newblock {SOLOMON: Seeking the Truth Via Copying Detection}.
\newblock {\em PVLDB}, 3(2):1617--1620, 2010.

\bibitem{Dong:2010}
X.~L. Dong, L.~Berti-Equille, Y.~Hu, and D.~Srivastava.
\newblock {Global Detection of Complex Copying Relationships Between Sources}.
\newblock {\em Proc. VLDB Endow.}, 3(1-2):1358--1369, 2010.

\bibitem{DongBS09}
X.~L. Dong, L.~Berti-Equille, and D.~Srivastava.
\newblock Integrating conflicting data: The role of source dependence.
\newblock {\em PVLDB}, 2(1):550--561, 2009.

\bibitem{DongBS09a}
X.~L. Dong, L.~Berti-Equille, and D.~Srivastava.
\newblock {Truth Discovery and Copying Detection in a Dynamic World}.
\newblock {\em PVLDB}, 2(1):562--573, 2009.

\bibitem{vldb14}
X.~L. Dong, E.~Gabrilovich, G.~Heitz, W.~Horn, K.~Murphy, S.~Sun, and W.~Zhang.
\newblock { From Data Fusion to Knowledge Fusion}.
\newblock In {\em VLDB}, 2014.

\bibitem{GallandAMS10}
A.~Galland, S.~Abiteboul, A.~Marian, and P.~Senellart.
\newblock {Corroborating Information from Disagreeing Views}.
\newblock In {\em WSDM}, pages 131--140, 2010.

\bibitem{GoasdoueKKLMZ13}
F.~Goasdou{\'e}, K.~Karanasos, Y.~Katsis, J.~Leblay, I.~Manolescu, and
  S.~Zampetakis.
\newblock {Fact Checking and Analyzing the Web}.
\newblock In {\em SIGMOD}, pages 997--1000, 2013.

\bibitem{sigmod14}
Q.~Li, Y.~Li, J.~Gao, B.~Zhao, W.~Fan, and J.~Han.
\newblock { Resolving Conflicts in Heterogeneous Data by Truth Discovery and
  Source Reliability Estimation }.
\newblock In {\em SIGMOD}, 2014.

\bibitem{LiDLMS12}
X.~Li, X.~L. Dong, K.~Lyons, W.~Meng, and D.~Srivastava.
\newblock {Truth Finding on the Deep Web: Is the Problem Solved?}
\newblock {\em PVLDB}, 6(2):97--108, 2012.

\bibitem{Pasternack:2010}
J.~Pasternack and D.~Roth.
\newblock Knowing what to believe (when you already know something).
\newblock In {\em COLING '10}, pages 877--885, 2010.

\bibitem{PasternackR13}
J.~Pasternack and D.~Roth.
\newblock {Latent Credibility Analysis}.
\newblock In {\em WWW}, pages 1009--1020, 2013.

\bibitem{Pochampally2014}
R.~Pochampally, A.~D. Sarma, X.~L. Dong, A.~Meliou, and D.~Srivastava.
\newblock {Fusing Data with Correlations}.
\newblock In {\em SIGMOD}, 2014.

\bibitem{VydiswaranZR11}
V.~G.~V. Vydiswaran, C.~Zhai, and D.~Roth.
\newblock Content-driven trust propagation framework.
\newblock In {\em KDD}, pages 974--982. ACM, 2011.

\bibitem{WangKLA12}
D.~Wang, L.~M. Kaplan, H.~K. Le, and T.~F. Abdelzaher.
\newblock {On Truth Discovery in Social Sensing: a Maximum Likelihood
  Estimation Approach}.
\newblock In {\em IPSN}, pages 233--244, 2012.

\bibitem{YinHY08}
X.~Yin, J.~Han, and P.~S. Yu.
\newblock {Truth Discovery with Multiple Conflicting Information Providers on
  the Web}.
\newblock {\em TKDE}, 20(6):796--808, 2008.

\bibitem{ZhaoRGH12}
B.~Zhao, B.~I.~P. Rubinstein, J.~Gemmell, and J.~Han.
\newblock {A Bayesian Approach to Discovering Truth from Conflicting Sources
  for Data Integration}.
\newblock {\em PVLDB}, 5(6):550--561, 2012.

\end{thebibliography}
 \end{scriptsize}
\end{document}